\documentclass[prd,twocolumn,preprintnumbers]{revtex4-1}
\usepackage{amsthm}
\usepackage{mathtools}
\usepackage{physics}
\usepackage{xcolor}
\usepackage{graphicx}
\usepackage[left=23mm,right=13mm,top=35mm,columnsep=15pt]{geometry} 
\usepackage{adjustbox}
\usepackage{placeins}
\usepackage[T1]{fontenc}
\usepackage{lipsum}
\usepackage[utf8]{inputenc}
\usepackage{csquotes}
\usepackage[english]{babel}
\usepackage{amsmath}
\usepackage{titlesec}
\usepackage{soul}
\usepackage{soulutf8}
\sethlcolor{white}
\titlespacing{\section}{0pt}{*5}{*2}
\titlespacing{\subsection}{0pt}{*3}{*2}
\usepackage{enumitem}
\usepackage{hyperref}
\hypersetup{pdftitle={Article}, pdfauthor={Author}}

\usepackage{array,multirow,xcolor,colortbl}
\usepackage{adjustbox}
\definecolor{amber}{HTML}{FAD7A0}
\definecolor{redDark}{HTML}{EA9999} 
\definecolor{greenMed}{HTML}{B7DDBF}
\newcolumntype{C}[1]{>{\centering\arraybackslash}m{#1}}

\begin{document}

\bibliographystyle{apsrev4-1}

\title{MaxWave: Rapid maximum likelihood
\\ wavelet reconstruction of non-Gaussian features in gravitational wave data}
\author{Sudhi Mathur}
    \affiliation{eXtreme Gravity Institute, Department of Physics, Montana State University, Bozeman, Montana 59717 USA}

\author{Neil J. Cornish}
    \affiliation{eXtreme Gravity Institute, Department of Physics, Montana State University, Bozeman, Montana 59717 USA}
    
\date{July 23, 2025}

\begin{abstract}
\noindent Advancements in the sensitivity of gravitational wave detectors have increased the detection rate of transient astrophysical signals. We improve the existing \textit{BayesWave} initialization algorithm and present a rapid, low-latency approximate maximum likelihood solution for reconstructing non-Gaussian features. We include three enhancements: (1) using a modified wavelet basis to eliminate redundant inner product calculations; (2) shifting from traditional time-frequency-quality factor wavelet transforms to time-frequency-time extent transforms to optimize wavelet subtractions; (3) implementing a downsampled heterodyned wavelet transform to accelerate initial calculations. Our model can be used to denoise long-duration signals, which include the stochastic gravitational wave background from numerous unresolved sources and continuous wave signals from isolated sources such as rotating neutron stars. \hl{Through our model, we can also supplement machine learning applications that use spectrographic training data to classify and understand glitches by providing nonwhitened, time and frequency domain reconstructions of any glitch.} 
\end{abstract}


\maketitle

\section{INTRODUCTION} \label{sec:Introduction}

The improved sensitivity \cite{LVK, AdLIGO-O4-2025, Soni_2025, Di_Pace_2021} of gravitational wave detectors \cite{LVK, AdLIGO-2015, Abbott_2009, Acernese_2008} has increased the detection of both astrophysical signals and transient, non-Gaussian noise events called ``glitches''. Since most gravitational wave noise modeling and data analysis algorithms assume Gaussian and stationary noise \cite{Abbott_2020}, such non-Gaussian features introduce outliers and disruptions. Glitches mimic or mask true astrophysical signals, complicating both the detection \cite{Davis_2022, Payne} and parameter estimation \cite{Sudarshan-2024, Sophie-2022} of transient signals. Both glitches and transient signals can dominate weak, long-duration signals such as the stochastic gravitational wave background from numerous unresolved sources and continuous wave signals from isolated sources such as rotating neutron stars. These non-Gaussian artifacts can create abrupt, large deviations in the data, making it difficult to track the frequency evolution of continuous signals and provide accurate astrophysical inference \cite{Isotropic-GWB-search, Anisotropic-GWB-search, CW-search, CW-search-TechReport}. While algorithms like \textit{BayesWave} \cite{Cornish_2015} can identify and subtract glitches and transient signals for stochastic and continuous wave searches, we must improve their efficiency to keep up with the rapidly growing number of non-Gaussian features in future LIGO-Virgo-KAGRA Collaboration (LVK) observing runs \cite{LVK}.

Additionally, programs like Gravity Spy \cite{Gravity_Spy_2017} use machine learning (ML) to classify, understand, and eventually prevent glitches before they interfere with gravitational wave searches, preserving the Gaussian nature of noise assumed for most analyses. Glitches can last from milliseconds to seconds and are extremely difficult to classify due to their highly diverse sources, characteristics, and morphological complexity. They can originate from a wide range of instrumental sources such as laser fluctuations, mirror vibrations, electronic noise in control systems, slight misalignments of the interferometer components, control system's response to drifts in alignment, and environmental factors such as temperature fluctuations, seismic activity, and wind. Most glitches do not have well-understood origins, and their characteristics can change over time as the detector undergoes upgrades or maintenance \cite{Abbott_2020}. Gravity Spy aims to better understand instrumental noise by crowd-sourcing glitch classification and using convolutional neural networks to automate the process \cite{Gravity_Spy_2017}. However, the program only uses time-frequency-quality factor scans (TFQ maps/ Qscans) of various glitches as their training set. Glitch features extracted directly from the Qscans are subject to distortion due to the data whitening process, Gaussian noise, \hl{and the choice of Q parameters.} 

In this paper, we present a faster, low-latency model for reconstructing non-Gaussian features in gravitational wave data. We use our model to speed up glitch identification and subtraction. \hl{As different time-frequency methods intrinsically give different representations of glitches,} we render extra information about glitches' time and frequency domain attributes by \hl{extracting waveform features, separating the dominant non-Gaussian component from the Gaussian background, and reconstructing the glitch in the original, nonwhitened basis.} This extra information can significantly aid ML algorithms, which currently use Qscans to understand instrument noise.

Our model is an improved version of the BayesWave \textit{FastStart} approximate maximum likelihood algorithm \cite{Neil-BWstart-2021}, which extracts non-Gaussian features from ground-based detectors. The existing algorithm reduces the convergence time for \textit{BayesWave} noise model \cite{Cornish_2015} over its large parameter space by starting the sampler's reversible-jump Markov Chain Monte Carlo (RJMCMC) chains near a good initial solution \cite{Neil-BWstart-2021}. This is crucial because the \textit{BayesWave} noise model uses hundreds of parameters to describe spline control points, Lorentzian lines, and glitch model wavelets in gravitational wave data. Without a good initial solution for this extensive parameter space, the sampler may require thousands of iterations to converge or reach an equilibrium posterior probability distribution. Through modifications described in this paper, we make the calculation of the glitch model significantly more efficient.

Currently, the approximate maximum likelihood solution of the BayesWave \textit{FastStart}'s Glitch model \cite{Neil-BWstart-2021} is computed by whitening the data using an initial power spectral density estimate. The whitened data is then wavelet transformed at a geometric sequence of quality factor values using continuous Morlet-Gabor wavelets on a time and frequency grid. The wavelet transform creates an overcomplete TFQ map dense enough to capture the signal efficiently with the least number of wavelets \cite{Chatterji-2004}. The loudest pixel (or region with the highest power) in the TFQ map is identified and used to reconstruct a wavelet, which is then subtracted from the whitened data. Since continuous, nonorthogonal wavelets overlap with their neighbors, the TFQ map is recomputed in the frequencies surrounding the subtracted wavelet. Wavelets corresponding to the brightest pixel are subtracted, and the TFQ map is recomputed iteratively until the signal-to-noise ratio (SNR) falls below a certain threshold. \hl{The reconstructed wavelets are clustered by proximity in grid space, after which a combined approximate maximum likelihood solution identifies the minimum number of wavelets that best represents the glitch or signal.}

\begin{figure}
    \centering
    \includegraphics[width=8.5cm]{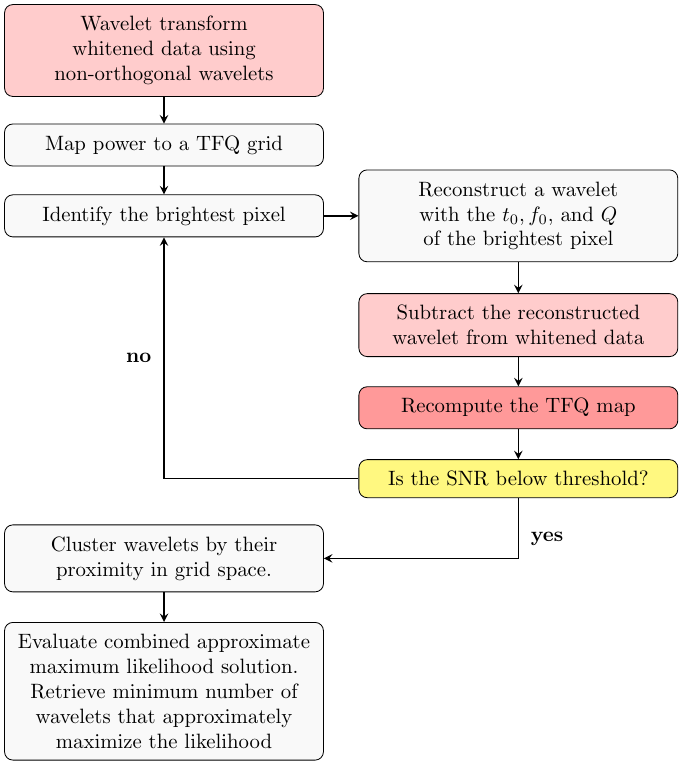}
    \caption{Traditional model \cite{Neil-BWstart-2021} for reconstructing noise transients using an approximate, iterative maximum likelihood solution. This model has several limitations including the costs of calculating the original wavelet transform using an overcomplete basis, recomputing noise-weighted inner products for updated TFQ wavelet transforms after each subtraction, and subtracting a large number of pixels as power heavily smears across different Q layers. The computational bottlenecks of the algorithm are highlighted in shades of red.}
    \label{fig:Glitchmodel_flowchart}
\end{figure}

\hl{In this paper, we highlight and improve major computational limitations of \textit{BayesWave}'s approximate maximum likelihood solution (Fig.} \ref{fig:Glitchmodel_flowchart}). Firstly, recomputing parts of the TFQ map is inefficient and computationally intensive as it involves recalculating noise-weighted inner products of the wavelet basis with the subtracted wavelet. We propose to remove the noise spectrum dependence of our wavelet inner products by altering our wavelet basis. This modified or ``white'' wavelet basis allows us to switch to analytic calculations for the wavelet inner products and eliminate the computational cost of recomputing the wavelet transform after each subtraction. Secondly, the time-frequency resolution changes between Q layers, smudging the spread of power and increasing the number of pixels we need to subtract as we move further away from the Q layer of the brightest wavelet. We switch to a time-frequency-time extent (TF$\tau$) map to reduce the number of pixel subtractions and analytical calculations. In contrast to TFQ maps, for the TF$\tau$ maps, the power of the brightest wavelet is well-localized in all the $\tau$ layers, reducing the number of pixels we need to subtract and the number of iterations required. Finally, to speed up the initial wavelet transform used to create the TF$\tau$ map, we employ downsampled, heterodyned wavelet transform calculations, which base-band segments of the whitened data and interpolate it with wavelets at a fixed low frequency. Heterodyning allows us to resort to a coarser time resolution in the TF$\tau$ map and significantly reduce the number of inverse fast Fourier transforms (IFFTs) required in the initial wavelet transform. \\

\hl{Our modifications provide considerable gains:}
\begin{itemize}
    \item \hl{Modifying the wavelet basis eliminates the need to recompute the wavelet transform after each subtraction, providing the most significant speedup and making real-time analysis feasible;} 
    \item \hl{Switching to $\tau$ space better localizes wavelet power, reducing the number of pixel subtractions;}
    \item \hl{Downsampling and heterodyning accelerate the initial wavelet transform and further reduce runtime.}
\end{itemize}

\hl{We compare our model MaxWave's performance with the following algorithms:}
\begin{enumerate}
    \item \hl{BayesWave \textit{FastStart}'s Glitch model} \cite{Neil-BWstart-2021}\hl{, which MaxWave builds upon and improves.}
    \item[] \hl{\textit{FastStart} suffers from several computational bottlenecks that make real-time multidetector analysis infeasible (Fig.} \ref{fig:Glitchmodel_flowchart}). \hl{Moreover, it reconstructs only whitened waveforms on a fixed TFQ grid and provides no error envelopes. In contrast, MaxWave introduces several novel advantages. It can:}
    \begin{itemize}
        \item \hl{render nonwhitened reconstructions using our modified wavelet basis;}
        \item \hl{produce \textit{BayesWave}-like error envelopes derived from the Fisher information matrix;}
        \item \hl{perturb and refine the solution beyond the fixed TF$\tau$ grid.} 
    \end{itemize} 
    \item[] \hl{MaxWave's computational efficiency allows it to run real-time on multidetector networks and generate large training datasets for glitch classification. Considering these aspects, we compare BayesWave \textit{FastStart} and MaxWave in terms of operations cost, runtime, and added features of the algorithm.}
    \item \hl{The \textit{BayesWave} RJMCMC} \cite{Cornish_2015}, \hl{which MaxWave is designed to complement.}
    \item[] \hl{While \textit{BayesWave} achieves high-fidelity whitened reconstructions, it is computationally intensive, requiring hours of runtime and heavy likelihood evaluations. As MaxWave is capable of rendering high quality reconstructions, especially for higher SNRs and binary mass ratios, we test it against the RJMCMC's superior reconstruction fidelity.}
\end{enumerate} 

\begin{figure}[h]
    \centering
    \includegraphics[width=8.5cm]{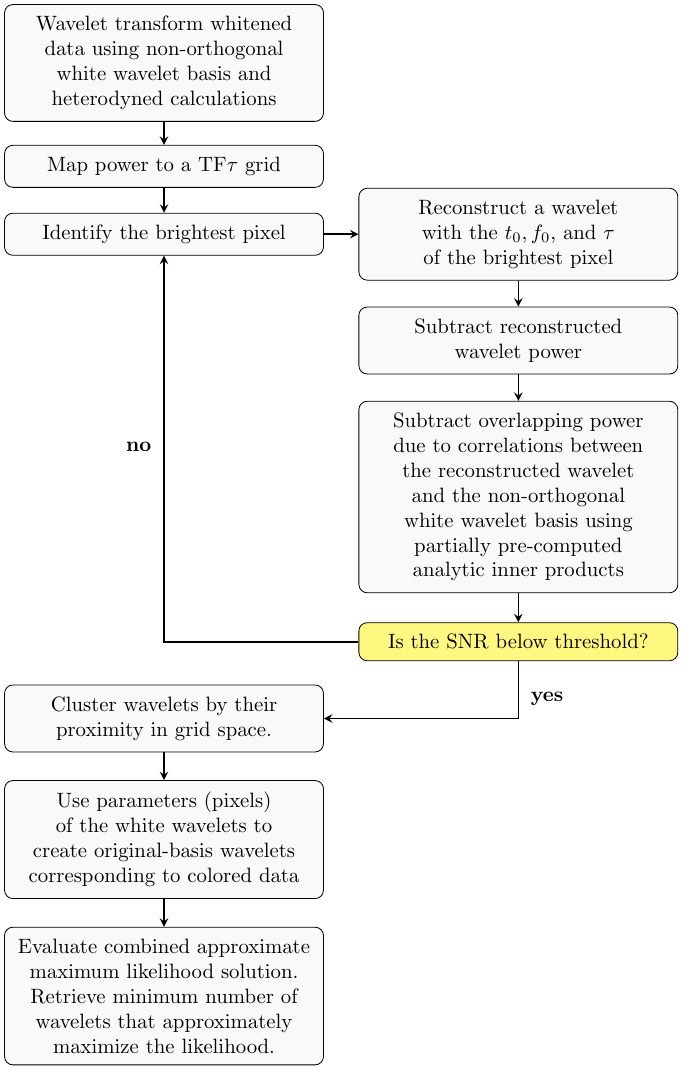}
    \caption{Modified model for constructing the approximate, iterative maximum likelihood solution. The computational cost of the original wavelet transform is reduced by base-banding whitened data and using downsampled heterodyned calculations. The hefty cost of recomputing TF$\tau$ transforms after each subtraction is entirely avoided by shifting to analytic wavelet inner products using white wavelets. The number of iterations or pixel subtractions required is also significantly reduced for the TF$\tau$ maps as the power of the brightest wavelet is well-localized in all the $\tau$ layers. Wavelet parameters retrieved can be used to reconstruct the signal corresponding to the approximate maximum likelihood solution.}
    \label{fig:MaxWave_flowchart}
\end{figure}

\hl{In the following sections, we present the formulation, implementation, and performance of MaxWave. Sec.} \ref{sec:Methods} \hl{describes our methodology. We construct a nonorthogonal modified wavelet basis with analytical inner products, derive the maximum-likelihood solution, and downsample and heterodyne the initial wavelet transform. We estimate false detection rates, quantify error envelopes, and explore perturbative refinements. Sec.} \ref{sec:Results} \hl{presents our results. We compare MaxWave with the BayesWave \textit{FastStart} algorithm} \cite{Neil-BWstart-2021} \hl{in terms of the output waveform, selection bias, and computational efficiency, and with the \textit{BayesWave} RJMCMC} \cite{Cornish_2015} \hl{in terms of reconstruction accuracy across various SNRs and binary mass ratios. We also present nonwhitened time and frequency domain reconstructions for various glitch types. Sec.} \ref{sec:Conclusion} \hl{summarizes our conclusions. Sec.} \ref{sec:Future Directions and Scope} \hl{discusses broader applications, including denoising long-duration signals, accelerating \textit{BayesWave} convergence and likelihood calculations, extending to low-latency burst searches, and complementing machine learning glitch classifiers.}

\section{METHODS} \label{sec:Methods}

With our proposed improvements (Fig. \ref{fig:MaxWave_flowchart}) to the traditional glitch model, we transform whitened data using white nonorthogonal wavelets (further described in Secs. \ref{subsec:Glitch model and nonorthogonal wavelets} and \ref{subsec:Modified wavelet basis}) and employ downsampled, heterodyned wavelet transform calculations to compute an initial TF$\tau$ map (Sec. \ref{subsec:Downsampled, Hereodyned Wavelet Transforms}). \hl{For any transient, one $\tau$ layer typically contains the maximum-power pixel on the TF$\tau$ map, while the full map preserves information across all temporal scales.} We iteratively find this brightest pixel on the TF$\tau$ map, reconstruct a wavelet with parameters corresponding to the brightest pixel, and subtract the reconstructed wavelet. We use analytical inner products between white wavelets that no longer depend on the noise spectrum to account for correlations between the subtracted wavelet and the wavelet basis and simultaneously subtract any overlapping power (Sec. \ref{subsec:Analytical Maximum Likelihood solution}). \hl{After clustering the identified white wavelets by grid space proximity, we use their parameters (or pixels) to reconstruct wavelets in the original basis --- corresponding to colored data --- and evaluate the combined approximate maximum likelihood solution.} \hl{In Secs.} \ref{subsec:False Detection Rate} to \ref{subsec:Iterations}, \hl{we estimate false detection rates, quantify error envelopes, and explore perturbative refinements.}

\hl{Although both the TFQ transform used in the traditional model and the TF$\tau$ transform used in our modified model are overcomplete, the former is logarithmically spaced while the latter is linearly spaced in frequency. Thus, in the discreet wavelet world, the closest analog to the TFQ transform is the dyadic standard wavelet decomposition --- characterized by a frequency resolution that doubles and a time resolution that halves between successive scales --- while that to the TF$\tau$ transform is the wavelet wave packet transform or binary decomposition (Fig. 3 in} \cite{SKlimenko_2004}). \hl{The TF$\tau$ transform is structurally closer to a short-time Fourier transform with a Gaussian window function, where each $\tau$-layer has a fixed temporal resolution.}

\subsection{Nonorthogonal wavelets and glitch model} \label{subsec:Glitch model and nonorthogonal wavelets}

To efficiently capture signal or glitch power, we need a dense, overcomplete TF$\tau$ grid made by transforming nonorthogonal wavelets \cite{Neil-BWstart-2021}. We employ nonorthogonal, overcomplete Morlet-Gabor wavelets \cite{Cornish_2015} that maximize the approximate likelihood using a minimum number of wavelets.

We reconstruct the signal/glitch model that has the form $h = \alpha_i \psi_i$, where $\alpha_i$ are complex coefficients and $\psi_i$ represent the nonorthogonal wavelets expressed in the time-domain as 
\begin{equation}\label{eq:wavelet}
    \Psi (t; \Vec{\lambda}) = A e^{-(t-t_0)^2/\tau^2} \cos{(2 \pi f_0 (t-t_0) + \phi_0)}, 
\end{equation}
\noindent
where the wavelet parameters are $\vec{\lambda} \rightarrow (A, t_0, f_0,$ Q or $\tau, \phi_0)$ such that $A$ is the wavelet amplitude, $t_0$ is the central time of the wavelet, $f_0$ is the frequency and $\phi_0$ is the phase at $t = t_0$, Q is the wavelet quality factor (i.e. number of cycles of the wavelet over one e-fold of the Gaussian envelope), and time extent $\tau = Q/2\pi f_0$. Fourier transforming the wavelet to the frequency domain, we get
\begin{equation}
    \begin{aligned}
    \Psi (f; \vec{\lambda}) & = \frac{\sqrt{\pi} A \tau}{2} e^{-\pi^2 \tau^2 (f-f_0)^2} e^{-2 \pi i f t_0}(e^{i \phi_0} \\
    & + e^{-i\phi_0} e^{-Q^2 f/f_0}).
    \end{aligned}
\end{equation}

In time-frequency space, Morlet-Gabor wavelets are ellipses with semiminor and semimajor axes of length $\Delta t = \tau$ and $\Delta f = 1/(2 \pi \tau)$ aligned with the time and frequency directions respectively, and area $\Delta t \Delta f = 1/2 \pi$ \cite{Chatterji-2004, Cornish_2015}. \hl{When moving across time–frequency layers, the shape of these pixels depends on whether the transform is parametrized by Q or by $\tau$. For TFQ maps, each layer has a fixed number of oscillations within the Gaussian envelope with $\Delta t = Q/2 \pi f_0$ and $\Delta f = f_0/Q$. Increasing Q makes each pixel longer in time and narrower in frequency, but the change is frequency dependent. At low frequencies the $\Delta t$ grows rapidly while at high frequencies it shrinks, distorting pixel shapes across the frequency axis. The inverse is true for $\Delta f$. As a result, the pixel aspect ratio varies strongly with frequency, and power from a single, fixed-duration signal smears across the multiple Q layers. In contrast, in a TF$\tau$ map, increasing $\tau$ also makes pixels longer in time and narrower in frequency, but this scaling occurs uniformly across all frequencies. Every frequency band shares the same time and frequency resolution, so the pixel shape remains constant. Consequently, signals of a fixed duration align coherently within a single $\tau$ layer, yielding well-localized power rather than frequency-dependent smearing. We switch to a TF$\tau$ map to take advantage of this localization of power and reduce the number of pixel subtractions for each bright wavelet identified.}

\subsection{Modified wavelet basis}\label{subsec:Modified wavelet basis}
We modify the wavelet basis to define ``white'' wavelets such that their inner product does not depend on the noise spectrum. Effectively, we do this by switching the order of whitening the data and representing it as a collection of wavelets. 

\hl{Let $W$ denote the whitening operator, which Fourier transforms a time-series $d$ to frequency domain with operator $\mathcal{F}$ and filters it with the square-rooted inverse of the power spectral density $1/\sqrt{S(f)}$, such that whitened data $d_w = W[d] = \mathcal{F}^{-1} \{\mathcal{F} \{d\}/\sqrt{S(f)}\}$. In the ``white'' wavelet basis, we first whiten both the data $d_w = W[d]$ and the basis $\psi_{w,i}  = W[\psi_i]$.} We then represent the whitened data as a sum of ``white'' wavelets $ d_w \rightarrow \beta_i \psi_{w,i}$, where the coefficients
\begin{equation}
    \beta_i = \sum_f (d_w^\ast \psi_{w,i} + d_w \psi_{w,i}^\ast),
\end{equation}
\noindent are independent of the power spectral density $S(f)$, rather than representing colored data as a sum of wavelets $ d \rightarrow \alpha_i \psi_i$ in the original basis such that 
\begin{equation}
    \alpha_i = \sum_f \frac{d^\ast \psi_i + d \psi_i^\ast}{S(f)}, 
\end{equation}
and then whitening the data using the noise spectrum. The whitening process and the summation do not commute. \hl{The noncommutation arises as the wavelets are nonorthogonal and $W$ is a nonlocal filtering operation. Applying $W$ to a nonorthogonal combination of wavelets introduces cross-terms among overlapping basis functions. Whitening the data first filters each wavelet individually, whereas whitening after summation mixes neighboring components through a convolution. Thus, when we whiten our data in the original basis, we get}
\begin{equation}
    W\left[ \sum_i \alpha_i \psi_i\right] \neq \sum_i \alpha_i W[\psi_i].
\end{equation}
\begin{figure}[h]
    \centering
    \includegraphics[width=8cm]{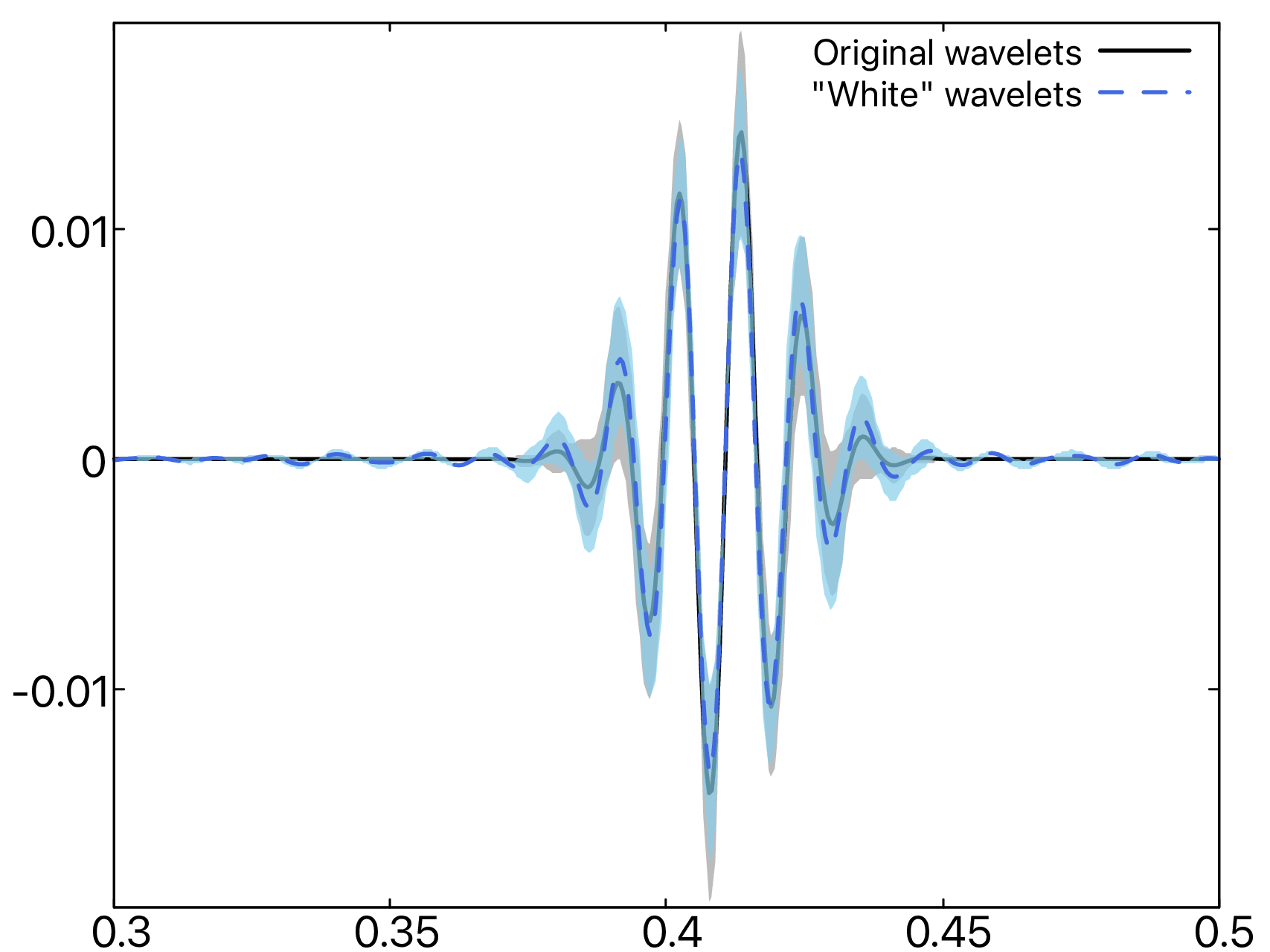}
    \includegraphics[width=8cm]{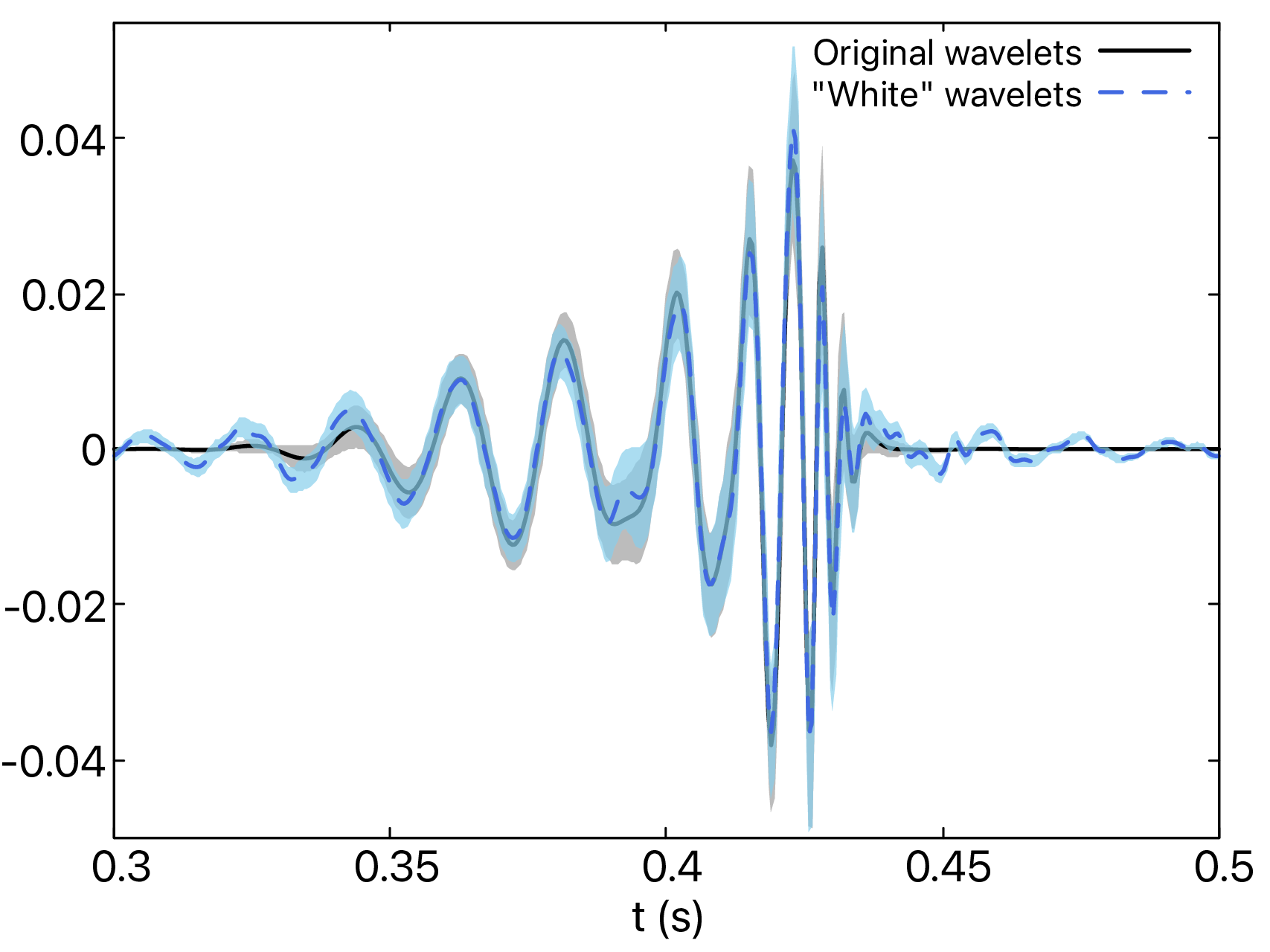}
    \caption{A single Morlet-Gabor wavelet in the original (gray waveform) and white basis (blue waveform) is plotted at the top. The bottom plot shows the gravitational wave signal GW150914 \cite{GW150914, GWTC-1, GWOSC_O1_O3} reconstruction using the original (gray waveform) and white (blue waveform) wavelet basis. Wavelets from the original basis are smooth in the whitened domain, whereas white wavelets show small-scale variations due to the whitening filter. \hl{However, we can use the $t_0, f_0$, and $\tau$ pixels from analytical subtractions in the white wavelet basis to reconstruct a physical signal in the original basis, and then compute the approximate maximum likelihood solution.}}
    \label{fig:Original-vs-White}
\end{figure}

\hl{Noncommutation of the whitening filter makes the white and original basis inherently different.} Fig. \ref{fig:Original-vs-White} (top) shows a single Morlet-Gabor wavelet in the original (gray waveform) and white wavelet basis (blue waveform). We use the Fisher information matrix to construct an error envelope around the waveform [Eq. (\ref{eq:variance})]. The wavelet bases render slightly different reconstructions of gravitational wave signals as the wavelet inner products in the bases are slightly different. The differences between whitened-domain reconstructions using original (gray waveform) and white wavelets (blue waveform) for the GW150914 event \cite{GW150914, GWTC-1, GWOSC_O1_O3} are also highlighted in Fig. \ref{fig:Original-vs-White} (bottom).  Wavelets from the original basis are smooth in the whitened domain, whereas white wavelets show small-scale variations due to the whitening filter. Thus, we only use white wavelets for pixel identification. \hl{Variations between the original and white wavelet basis do not introduce critical issues for pixel identification, as we apply the whitening filter across both the data and the wavelet basis, preserving the relative localization of power on the TF$\tau$ map.} \hl{We can then reconstruct the final signal by computing the combined approximate maximum likelihood solution for the original wavelets at pixels ($t_0, f_0$, and $\tau$) corresponding to the identified white wavelets.}

\subsection{Analytical maximum likelihood solution}\label{subsec:Analytical Maximum Likelihood solution}
\hl{We assume that the detector noise $n(t)$ is stationary, Gaussian, and characterized by $S(f)$. Given our assumption, the probability of observing the data $d(t) = n(t) + h(t)$ for a waveform model $h(t)$ is}
\begin{equation}
    p(d|h) = \frac{1}{Z}\exp\left[-\frac{1}{2}(d-h|d-h)\right],
    \label{eq:probability_density}
\end{equation}
\noindent \hl{such that the normalization constant $Z = \sqrt{(2 \pi)^N \text{det}C }$, where $C$ is the noise covariance matrix, $N$ is the number of data samples, and the inner product is defined in the original basis as}
\begin{equation}
    (a|b) = 2 \int_0^{\infty}{\frac{\tilde{a}^*(f) \tilde{b}(f) + \tilde{a}(f) \tilde{b}^*(f)}{S(f)}df}.
\end{equation}

\hl{We take the logarithm of Eq.} (\ref{eq:probability_density}) \hl{and calculate the log likelihood computed using our nonorthogonal filters in the original basis ($h = \alpha_i \psi_i$) as}
\begin{equation}
    \begin{aligned}
    \log{\mathrm{L}} & = -\frac{1}{2}(d|d) + (d|h) - \frac{1}{2}(h|h)\\
           & = -\frac{1}{2}(d|d) + \alpha_i (d|\psi_i) - \frac{1}{2}\alpha_i \alpha_j(\psi_i|\psi_j)\\
           & = -\frac{1}{2}(d|d) + \alpha_i N_i - \frac{1}{2}\alpha_i \alpha_j M_{ij},
    \end{aligned}
    \label{eq:log-likelihood}
\end{equation}
\noindent where the maximum likelihood solution is given by $\alpha_i = (M^{-1})_{ij} N_j$. \hl{Note that the log likelihood is evaluated in the original wavelet basis. The white wavelet basis serves only for initial TF$\tau$ transform, bright-pixel selection, and analytical subtractions}. Our calculation is similar to the F-statistic, a specific example of computing the likelihood using nonorthogonal filters \cite{F-statistic}. \hl{We invert the wavelet inner product matrix $M_{ij}$ using lower-upper (LU) decomposition with partial pivoting as implemented in the GNU Scientific Library (GSL)} \cite{GSL}. 

\hl{Unlike thresholded inverse continuous wavelet transforms (CWTs)} \cite{Torrence&Compo, WaveScan} \hl{that project all above-threshold pixels back into the time-domain simultaneously, MaxWave avoids the degeneracy that occurs in oversampled, nonorthogonal basis. For CWTs, $M_{ij}$ becomes nearly singular, and the solution is no longer unique. However, MaxWave iteratively subtracts the loudest pixel and removes overlapping power with neighboring pixels using analytical wavelet inner products. Thus, the bright pixels identified are further apart on a downsampled TF$\tau$ grid. Moreover, wavelets with significant overlap in time and frequency are clustered to form a single wavelet. Only a few distinct wavelets are identified with low overlap between any two wavelets. The resulting $M_{ij}$ matrix is sparse and well-conditioned --- with small off-diagonal elements --- and its LU factorization remains unique, stable and efficient.}

The analytic expression of the wavelet inner product in the white wavelet basis is given by
\begin{multline}
    M_{ij} = \sqrt{\frac{2 \tau_i \tau_j}{\tau_i^2 + \tau_j^2}} \cos{(\Delta \phi - 2 \pi \Delta t_0 \bar{f})} \\ \exp{-\frac{\Delta t_0^2+\pi^2 \tau_i^2 \tau_j^2 \Delta f_0^2}{\tau_i^2 + \tau_j^2}},
    \label{eq:Mij}
\end{multline}
\noindent where $t_0$, $f_0$, and $\tau$ are the central time, central frequency, and time extent of a wavelet, and $\Bar{f}=(f_{0,i} \tau_i^2 + f_{0,j} \tau_j^2)/(\tau_i^2+\tau_j^2)$, $\Delta \phi = \phi_i-\phi_j$,  $\Delta t_0 = t_{0,i}-t_{0,j}$, and $\Delta f_0 = f_{0,i}-f_{0,j}$ for wavelets $\psi_i$ and $\psi_j$. \hl{We generate the TF$\tau$ map by wavelet transforming whitened data in the white wavelet basis and use analytical inner products to iteratively subtract wavelet power.} We precompute and store the exponential part of the analytic inner products in Eq. (\ref{eq:Mij}), \hl{which is translationally invariant in time and frequency when the wavelet transform is parametrized by $\tau$ instead of Q}. \hl{After clustering the pixels identified from analytical subtractions in the white wavelet basis, we use them to reconstruct wavelets in the original basis and evaluate the approximate maximum likelihood solution.}

\subsection{Downsampled, heterodyned wavelet transforms}\label{subsec:Downsampled, Hereodyned Wavelet Transforms}

\begin{figure}
    \centering
    \includegraphics[width=8.5cm]{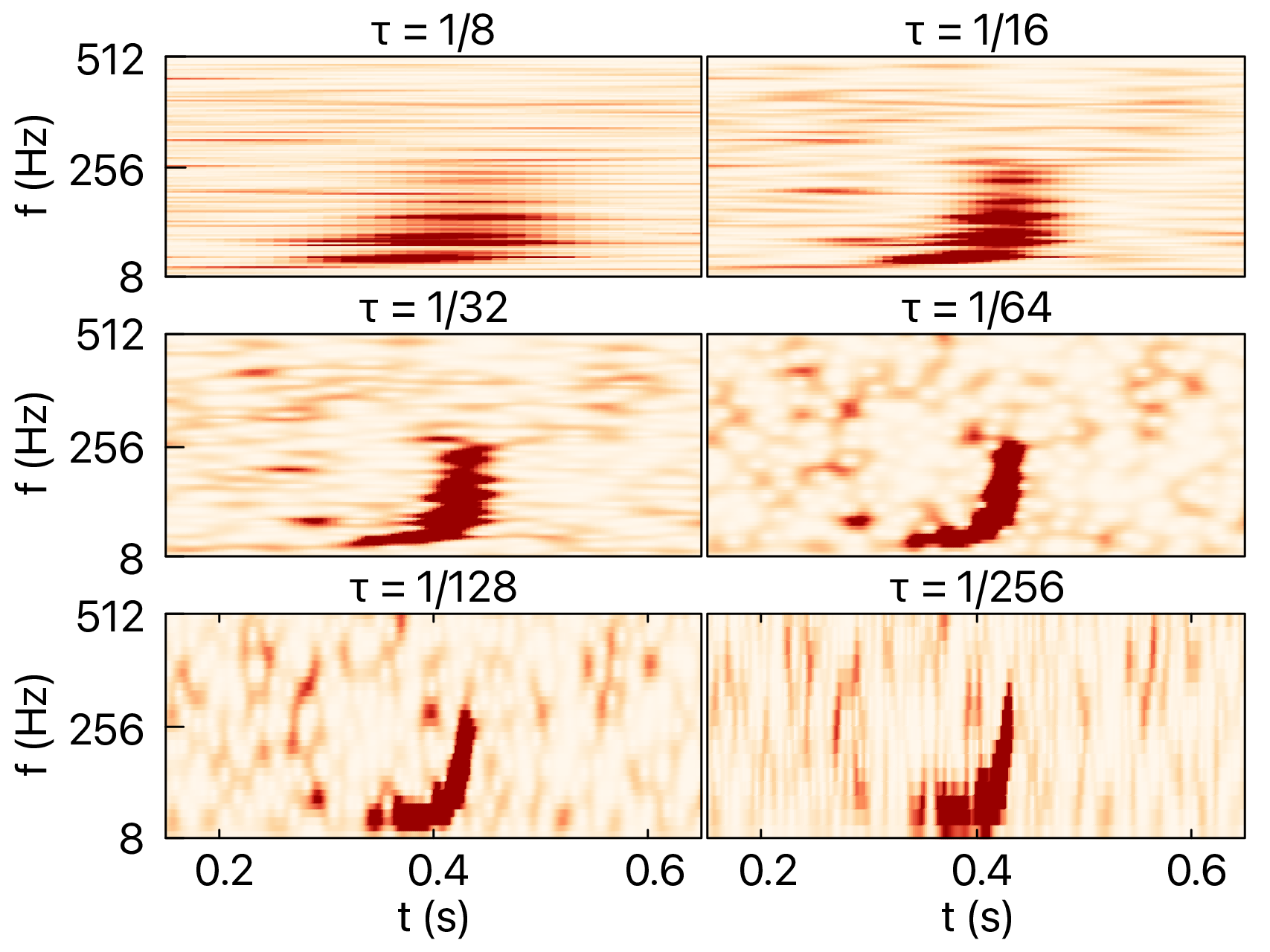}
    \caption{TF$\tau$ map for GW150914 signal \cite{GW150914, GWTC-1, GWOSC_O1_O3} in the LIGO Hanford detector calculated using wavelet transforms in the nonorthogonal, overcomplete, white wavelet basis. Each $\tau$ layer has a different time and frequency resolution. Pixel darkness is proportional to the signal-to-noise ratio of the signal. We boost the wavelet transform using $\tau$ dependent downsampling and heterodyned calculations that base-band segments of the whitened data and interpolate it with wavelets at a fixed low frequency.}
    \label{fig:TFtau_map}
\end{figure}

We speed up the initial wavelet transform of the whitened data in the Fourier domain using downsampling and heterodyned calculations. The TF$\tau$ map calculated at full resolution \hl{by placing wavelets at every $\Delta t$ and $\Delta f$,} is highly oversampled as the nonorthogonal wavelets significantly overlap in time. In our heterodyned transforms (Fig. \ref{fig:TFtau_map}), \hl{we adopt a coarser time resolution using the wavelet inner product in Eq.} (\ref{eq:Mij}), \hl{maximized over $\Delta \phi$ to choose the downsampled spacing. For small $\Delta t_0$, $\Delta f_0$ and $\Delta \tau$, we can expand}
\begin{equation}
    M_{ij} \approx 1 - \frac{\Delta \tau^2 + 2 \Delta t_0^2 +2 \pi^2 \Delta f_0^2 \bar{\tau}^4}{4 \bar{\tau}^2}, 
\end{equation}
\hl{where $\bar{\tau}^2 = (\tau_i^2 + \tau_j^2)/2$. For geometrically spaced $\tau$ layers in powers of $1/2$, we choose the downsampled time and frequency spacing}
\begin{equation}
    \Delta t' = \frac{\tau}{8}, \quad \quad \Delta f' = \frac{1}{8 \tau},
    \label{eq:downsample-spacing}
\end{equation}
\hl{such that the minimum overlaps or coverage for a wavelet precisely midway between two time, frequency, or time-extent layers are}
\begin{align}
     \Delta t_0: & \quad 1 - \frac{\Delta t_{0}^2}{2 \tau^2}  = 1 - \left(\frac{\Delta t'}{2}\right)^2 \frac{1}{2 \tau^2} = 0.998, \notag\\
     \Delta f_0: & \quad 1 - \frac{\pi^2 \Delta f_{0}^2 \tau^2}{2}  = 1 - \left(\frac{\Delta f'}{2}\right)^2 \frac{\pi^2 \tau^2}{2}  = 0.981, \notag\\
     \Delta \tau: & \quad 1 - \frac{\Delta \tau^2}{4 \tau^2} = 1 - \left( \frac{\Delta \ln{\tau}}{2} \right)^2 \frac{1}{4} = 0.970.
     \label{eq:overlaps}
\end{align}
\hl{Using our chosen $\Delta t'$ and $\Delta f'$, we get the number of time and frequency slices,}
\begin{equation}
    N_t = \frac{T_{\text{obs}}}{\Delta t'} = \frac{8T_{\text{obs}}}{\tau}, \quad \quad 
    N_f = \frac{f_{\text{Ny}}}{\Delta f'} = 8 \tau f_{\text{Ny}},
\end{equation} 
\noindent \hl{where $f_{\text{Ny}}$ is the Nyquist frequency and $T_{\text{obs}}$ is the observation time. This gives us a fixed grid space for every $\tau$ layer $N_f N_t = N_{\text{grid}} = 64 f_{\text{Ny}} T_{\text{obs}} = 32 N$ for time samples $N = f_s T_{\text{obs}}$ where sampling rate $f_s = 2 f_{\text{Ny}}$. Compared to the original excessively oversampled grid,} downsampling ensures that the heterodyned TF$\tau$ map would be closer to critically sampled in time and could reduce the $\Delta t$ overlap by up to two orders of magnitude depending on the $\tau$ layer.

We then base-band segments of the frequency-domain whitened data $\tilde{d}(f)$ to get $\tilde{d}'(f) = \tilde{d}(f-f_c)$, where \hl{$f_c = N_t/(4 T_{obs}) = 2/\tau$}. We interpolate the base-banded, downsampled data $\tilde{d}'(f)$ with wavelets at a fixed low frequency, $\tilde{w}(f)$, which significantly reduces the number of IFFTs required for the transform. The discrete wavelet transform $\Psi(n), n = 0,..,N_t$ for each frequency layer over a smaller range of time slices $N_t < N$, is given by
\begin{equation}
    \Psi(n) = \frac{1}{N_t} \sum^{N_t-1}_{k=0} \tilde{d}(f-f_c) \tilde{w}(f) e^{i 2 \pi n k/N_t}. 
\end{equation}
As the wavelets are translationally invariant in frequency, we recover the wavelet transform at all frequencies by accounting for a heterodyne phase shift $\Delta \phi' = 2 \pi (f-f_c) \Delta t'$. The computational cost of a single FFT for N samples of time is of order $O(N \log{(N)})$, and a wavelet transform costs twice as much as a FFT. Thus, the operations cost of our downsampled, heterodyned wavelet transform is $O(2 N_f N_t \log{(N_t)})$ for each $\tau$ layer.

It is worth noting that each downsampled $\tau$ layer in the TF$\tau$ map has a fixed resolution as $\Delta t'$ \hl{and $\Delta f'$ only depend on $\tau$ and not the frequency. This makes TF$\tau$ scans more amenable to heterodyning}. In contrast, downsampling the TFQ scan is much more tedious as $\tau$ (and consequently  $\Delta t'$ \hl{and $\Delta f'$}) changes for every Q layer as the frequency increases. If we used TFQ maps, every Q layer would have varying resolution, and we would have to heterodyne in frequency bands.

\subsection{False detection rate}\label{subsec:False Detection Rate}

\begin{figure}
    \centering
    \includegraphics[width=8cm]{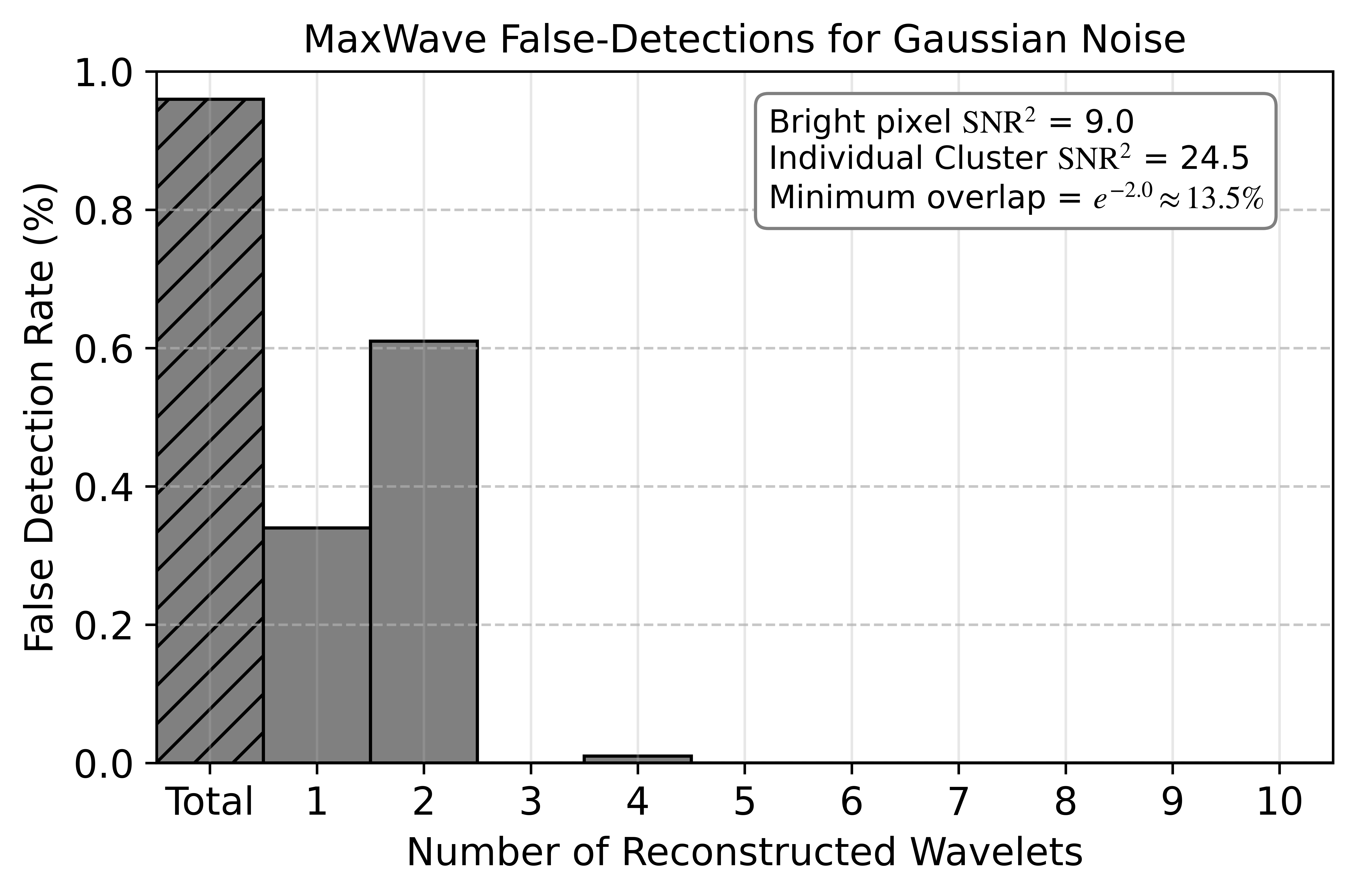}
    \caption{False detection rate for MaxWave recovery of ten-thousand, 4 second long Gaussian noise realizations. The histogram shows number of detections (first column) and the number of wavelets recovered (subsequent columns) as a fraction of the number of noise realizations. The false detection rate is less than 1 \%. Also note that  we intentionally suppress the detection of a single wavelet (second column) since glitches and signals are better modeled as clustered power.}
    \label{fig:False-Detection}
\end{figure}

We adjust our thresholds for identifying bright pixels such that the false alarm rate of the Morlet-Garbor wavelets with time-frequency area $\Delta t \Delta f  = 1/2 \pi $ is approximately 1\%. \hl{This is done using the $\text{SNR}^2$ of a Morlet-Gabor wavelet in white Gaussian noise, which follows a $\chi^2$ distribution with two degrees of freedom} \cite{Torrence&Compo}. \hl{We set the bright pixel threshold at $\rho_{\text{thresh}}^2{=}9$ $(\rho_{\text{thresh}}{=} 3)$, where the survival probability $P(\rho^2 > \rho_{\text{thresh}}^2) = e^{-\rho_{\text{thresh}}^2/2} = 1.\bar{1}\%$}. We also adjust our thresholds for the minimum overlap between wavelets ($e^{-2.0} \approx 13.5\%$) in a cluster and the minimum $\text{SNR}^2 = 24.5$ for an individual wavelet to form its own cluster. After these adjustments, MaxWave's false detection rate is less than 1\% for ten-thousand, 4 second long Gaussian noise realizations (Fig. \ref{fig:False-Detection}). By design, this is significantly higher than the false-detection rate for confident LVK detections \cite{LVK}, as we want to reconstruct low SNR glitches as well. Since glitches and signals are better modeled as clustered power, we intentionally suppress the detection of a single wavelet.

\subsection{Error envelopes}\label{subsec:Error Envelopes}

When fitting the signal $h(t)$ parametrized by the set of parameters $\vec{\lambda}$ to the data, \hl{detector noise and the finite, approximate parametrization of the model perturb} our solution as
\begin{equation}
    h'(t) = h(t) + \partial_k h(t) \Delta \lambda^k + \dots,
    \label{eq:perturb maximum likelihood solution}
\end{equation} 
where for Gaussian noise, covariance matrix $C$, and Fisher information matrix $\Gamma$, we get
\begin{equation}
{\rm E}[\Delta \lambda^k] = 0, \quad {\rm Var}[\Delta \lambda^k\Delta \lambda^l] = C^{kl} \simeq (\Gamma^{-1})^{kl} \, .
\end{equation} 
Thus, at each time sample, we have
\begin{equation}
{\rm E}[h'(t)] = h(t) \, ,
\end{equation} 
and 
\begin{equation}
{\rm Var}[h'(t)] = \partial_k h(t) \partial_l h(t) (\Gamma^{-1})^{kl} \, .
\label{eq:variance_unw}
\end{equation} 
The expressions are the same in the Fourier domain with $t \rightarrow f$.  \hl{The perturbations to our approximate maximum likelihood solution [Eq.} (\ref{eq:perturb maximum likelihood solution})] \hl{are quantified by the Fisher information matrix and determine the variance of the reconstructed waveform}. \hl{Note that the variance is evaluated directly on the reconstructed waveform in the time-domain and not in wavelet space. Thus, it already captures how uncertainties in the wavelet parameters translate into variations in the time-domain signal. This ensures that uncertainties associated with the inverse Fourier transform are reflected in the resulting error envelope.}  Now, whitening the signal before returning to the time domain, we get
\begin{multline}
{\rm Var}[h'(t)] = {\cal F}^{-1}\{\partial_k \tilde h(f)/ S(f)^{1/2}\} \\ {\cal F}^{-1}\{\partial_l \tilde h(f)/ S(f)^{1/2}\}   (\Gamma^{-1})^{kl} \, ,
\label{eq:variance}
\end{multline} 
where $ {\cal F}^{-1}$ denotes the inverse Fourier transform. As the Fisher matrix $\Gamma$ is real and symmetric, the expressions for the variance are quadratic forms and are guaranteed to be non-negative. Next, we compute the error envelopes for the Fourier-domain amplitude. The perturbed waveform can be written as
\begin{equation}
    \bar{h}(f) = \tilde{h}(f) + \partial_j \tilde{h}(f) \Delta \lambda^j + \frac{1}{2} \partial_k \partial_l \tilde{h}(f) \Delta \lambda^k \Delta \lambda^l \dots 
\end{equation}
The squared amplitude of the perturbed waveform is given by $\bar{A}^2(f) = \bar{h}(f)\bar{h}^*(f)$. Then $E[\bar{A}(f)]$ is $A(f) = \sqrt{\tilde{h}(f) \tilde{h}^*(f)}$ to leading order with higher order $(\Gamma^{-1})^{jk}$ terms. Using ${\rm Var}[\bar{A}(f)] = (\bar{A}(f) - E[\bar{A}(f)])^2$, we get
\begin{equation}
    \begin{aligned}
    {\rm Var}[\bar{A}(f)] & =  \frac{1}{4 \tilde{h}(f)\tilde{h}^*(f)} \left( \tilde{h}^*(f)\partial_j\tilde{h}(f) + \tilde{h}(f)\partial_j\tilde{h}^*(f) \right) \\ 
    & \hspace{0.4cm} \left( \tilde{h}^*(f)\partial_k\tilde{h}(f) + \tilde{h}(f)\partial_k\tilde{h}^*(f) \right) (\Gamma^{-1})^{jk} \\
    & = \frac{\partial_j A^2(f) \partial_k A^2(f)}{4 A^2(f)} (\Gamma^{-1})^{jk} \\
    & = \partial_j A(f) \partial_k A(f) (\Gamma^{-1})^{jk}.
    \end{aligned}
    \label{eq:variance_amp}
\end{equation}

We use Eqs. (\ref{eq:variance}) and (\ref{eq:variance_amp}) to construct error envelopes on our time and frequency domain approximate maximum likelihood wavelet reconstructions (Figs. \ref{fig:Original-vs-White}, \ref{fig:NH-C-vs-AH-W}, and \ref{fig:Glitches}).

\subsection{Refining the initial solution}\label{subsec:Iterations}

The approximate maximum likelihood solution we calculate is on a fixed TF$\tau$ grid. We can improve this initial solution by iterative refinements. Perturbing the solution with respect to the wavelet parameters $\vec{\lambda} \in (t_0, f_0, \tau, A, \phi_0)$ we get
\begin{equation}
    \tilde{h}(f) = \tilde{h}_{\text{max}}(f) + \Delta \tilde{h}(f) = \tilde{h}_{\text{max}}(f) + \partial_k h \Delta \lambda^k +  \dots
\end{equation}
We can write $\log{\mathrm{L}}$ in terms of the perturbation $\Delta \tilde{h}(f)$,
\begin{equation}
    \begin{aligned}
    \log{\mathrm{L}} & = -\frac{1}{2}((d- \tilde{h}_{\text{max}}(f)) - \Delta \tilde{h}(f))^2\\
           & = -\frac{1}{2} \bigg[ (d- \tilde{h}_{\text{max}}(f)| d- \tilde{h}_{\text{max}}(f)) \\
           &    - 2( \Delta \tilde{h}(f) | d- \tilde{h}_{\text{max}}(f)) + (\Delta \tilde{h}(f)| \Delta \tilde{h}(f)) \bigg].
    \end{aligned}
\end{equation}
Differentiating $\log{\mathrm{L}}$ with respect to $\Delta \lambda^k$, we get
\begin{equation}
    \frac{\partial \log{\mathrm{L}}}{\partial(\Delta\lambda^k)} = (\partial_k \tilde{h}(f)| d- \tilde{h}_{\text{max}}(f)) - (\partial_k \tilde{h}(f) | \partial_p \tilde{h}(f)) \Delta \lambda^p,
\end{equation}
where $(\partial_k \tilde{h}(f) | \partial_p \tilde{h}(f)) = \Gamma^{kp}$ gives the Fisher matrix elements. Thus, to further maximize the initial approximate $\log{\mathrm{L}}$, we can move our wavelet parameters by
\begin{equation}
    \Delta \lambda^p = (\partial_k \tilde{h}(f) |  d- \tilde{h}_{\text{max}}(f)) (\Gamma^{-1})^{kp}
    \label{Eqn:Iterations}.
\end{equation}
Note that maximizing $\log{\mathrm{L}}$ can lead us to fit more noise, especially for low SNR signals, and does not guarantee improvement in signal recovery. \hl{After obtaining the initial solution,} we perturb the $t_0, f_0, \tau$ wavelet parameters away from the fixed grid until we find a local maxima. We use $\log{\mathrm{L}}$ as a guide to explore the parameter space between our fixed grid points and fine tune our recovery. We iteratively shift the parameters by $\Delta \lambda^p$ until the consecutive increase in $\log{\mathrm{L}}$ drops below a certain threshold.

\section{RESULTS}\label{sec:Results}

We compare MaxWave with the BayesWave \textit{FastStart} algorithm \cite{Neil-BWstart-2021} in terms of the output waveform, selection bias, and computational efficiency (Sec. \ref{subsec:MW vs BW FastStart}), and with the \textit{BayesWave} RJMCMC \cite{Cornish_2015} in terms of reconstruction accuracy across various SNRs and binary mass ratios (Sec. \ref{subsec:MW vs BW}). We also present nonwhitened time and frequency domain reconstructions for various glitch types (Sec. \ref{subsec:TF reconstructions of Glitches}).

\subsection[\textit{FastStart}]{MaxWave versus BayesWave \textit{FastStart}}\label{subsec:MW vs BW FastStart}

\subsubsection{Reconstruction features}
\begin{figure}[b]
    \centering
    \includegraphics[width=8cm]{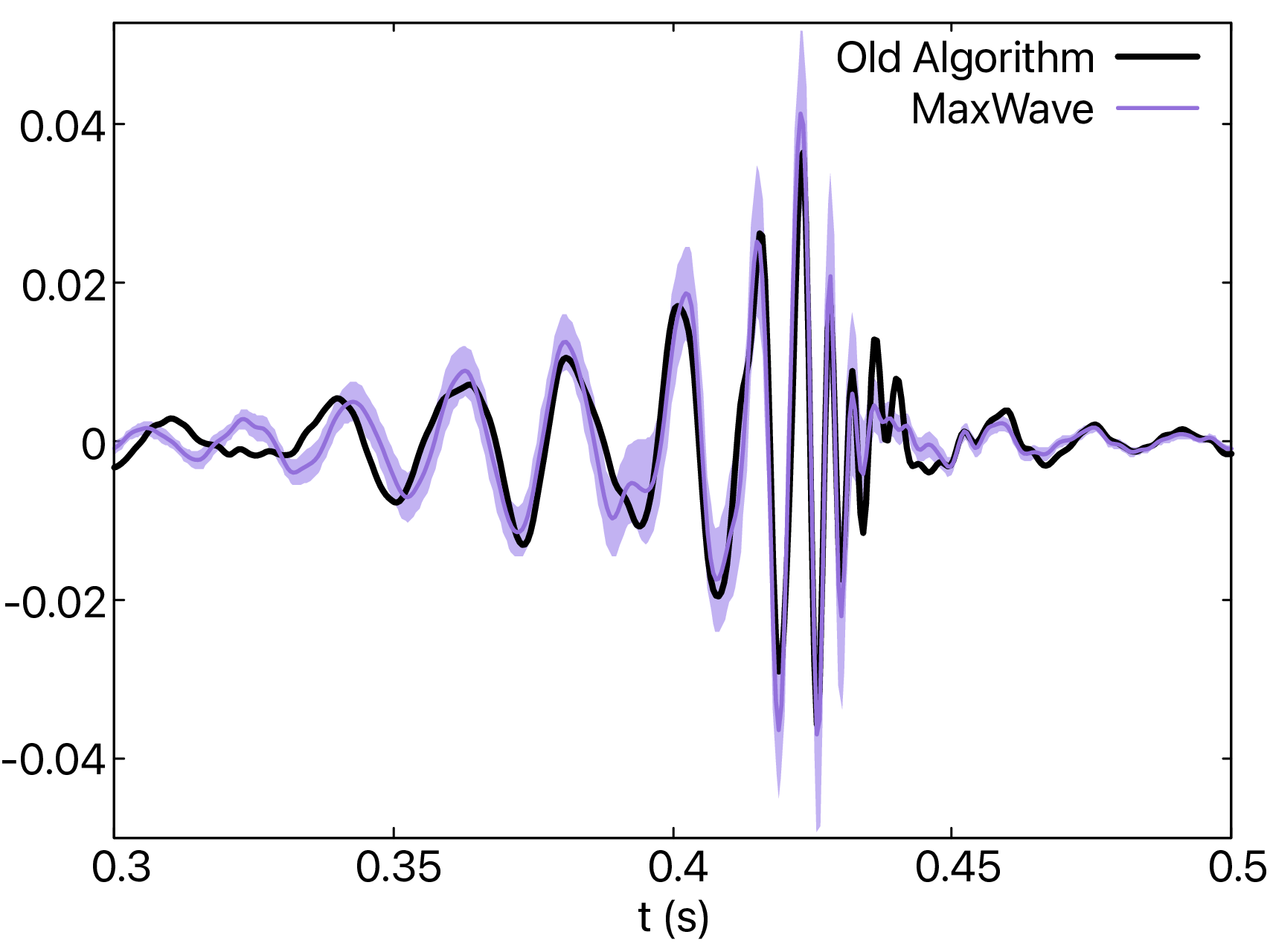}
    \caption{BayesWave \textit{FastStart} and MaxWave's time domain whitened reconstructions of gravitational wave signal GW150914 \hl{in the original basis}. We expect the wavelet parameters identified using transforms in the original (black waveform) and white (purple waveform) wavelet basis to be slightly different as the inner products of wavelets in the two bases are slightly different. Downsampling and heterodyning reduces the number of wavelets identified and clustered by MaxWave as compared to the old algorithm, where the $\Delta t$ overlap is extremely high. \hl{MaxWave's error envelopes are computed using Eq.} (\ref{eq:variance})}.
    \label{fig:NH-C-vs-AH-W}
\end{figure}
We compare the time-domain whitened reconstructions of the gravitational wave signal GW150914 \cite{GW150914, GWTC-1, GWOSC_O1_O3} produced by the old algorithm BayesWave \textit{FastStart} and MaxWave in Fig. \ref{fig:NH-C-vs-AH-W}. The old algorithm uses highly-oversampled wavelet transforms of whitened data and numerical recomputation of the transforms after each bright wavelet subtraction to produce the black waveform. MaxWave creates the purple waveform at minimal computational cost by using a single initial downsampled, heterodyned wavelet transform and analytical subtractions using partially precomputed, noise-independent wavelet inner products in the white wavelet basis. We expect the wavelet parameters identified using transforms in the original (black waveform) and white (purple waveform) wavelet basis to be slightly different, as the inner products of wavelets in the two bases are slightly different. Also note that downsampling and heterodyning reduces the overlap between nearby time pixels by upto two orders of magnitude depending on the $\tau$ layer. Thus, match values between nearby wavelets are lower, and fewer wavelets are identified and clustered by MaxWave (purple waveform) compared to the old algorithm (black waveform), where the $\Delta t$ overlap is extremely high. MaxWave's error envelopes are computed using Eq. (\ref{eq:variance}).

\begin{figure}
    \centering
    \includegraphics[width=8cm]{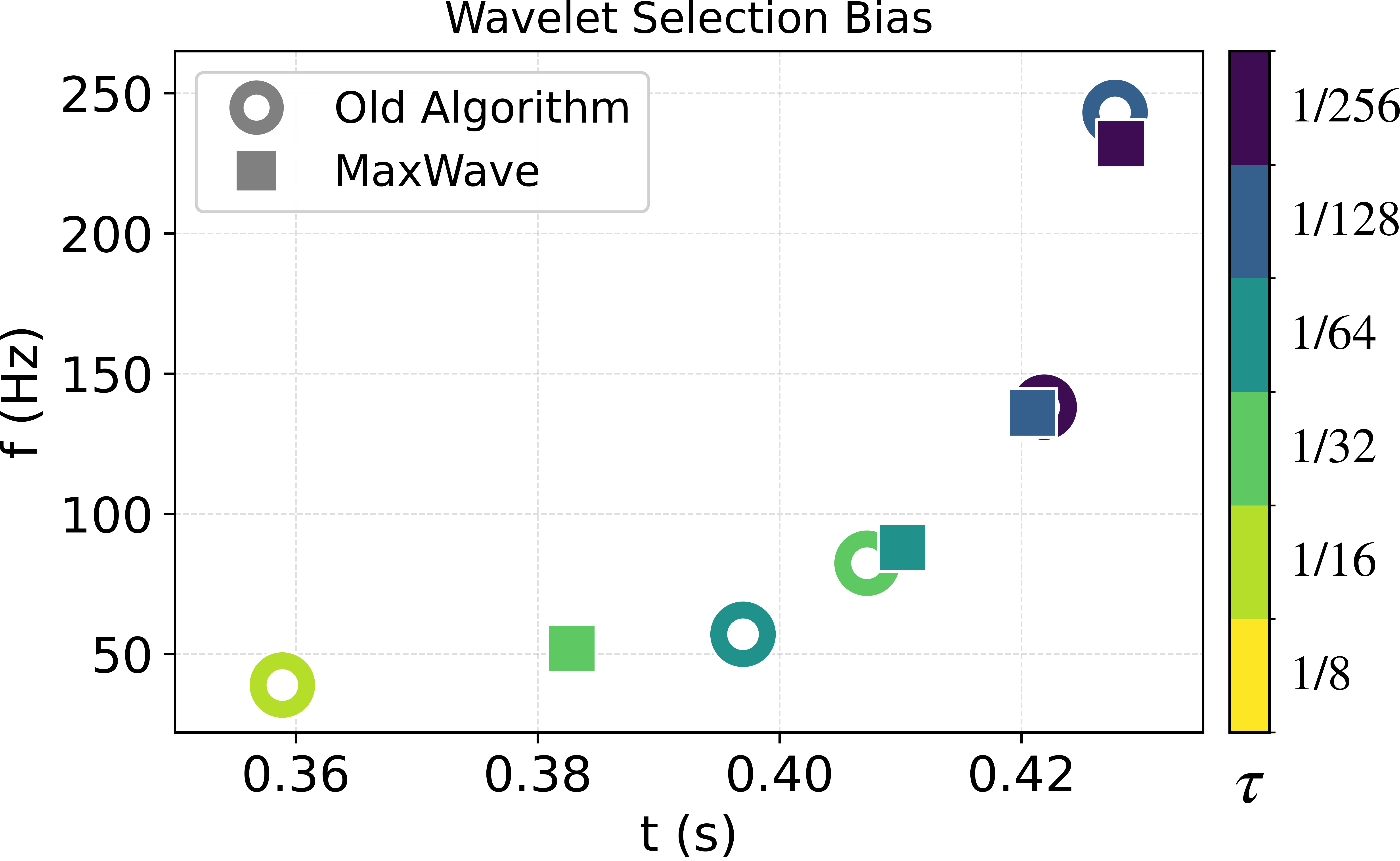}
    \caption{Comparing the $t_0, f_0$, and $\tau$ parameters selected by the old algorithm BayesWave \textit{FastStart} (rings), to those selected by MaxWave (squares) for the gravitational wave signal GW150914. \hl{The BayesWave \textit{FastStart} Q values are shown at the nearest MaxWave $\tau$ grid point using $\tau = Q / (2\pi f_0)$.} Note how the parameter values follow a three-dimensional time-frequency-$\tau$ evolution track.}
    \label{fig:Selection-Bias:All-modifications}
\end{figure}

Wavelets in the original and white wavelet basis have slightly different inner products that affect numerical or analytical subtractions and subsequent bright pixel identification. The wavelet parameters identified in our reconstructions are also affected by downsampling of the different $\tau$ layers used in our analysis. Fig. \ref{fig:Selection-Bias:All-modifications} compares wavelet identified by the old algorithm to those selected by MaxWave. \hl{BayesWave \textit{FastStart} Q values are shown at the nearest MaxWave $\tau$ grid point using the relation $\tau = Q / (2\pi f_0)$.} The parameter values selected are physically meaningful as they follow a three-dimensional time-frequency-$\tau$ evolution track as observed for binary black hole signals.

\begin{figure}
    \centering
    \includegraphics[width=8cm]{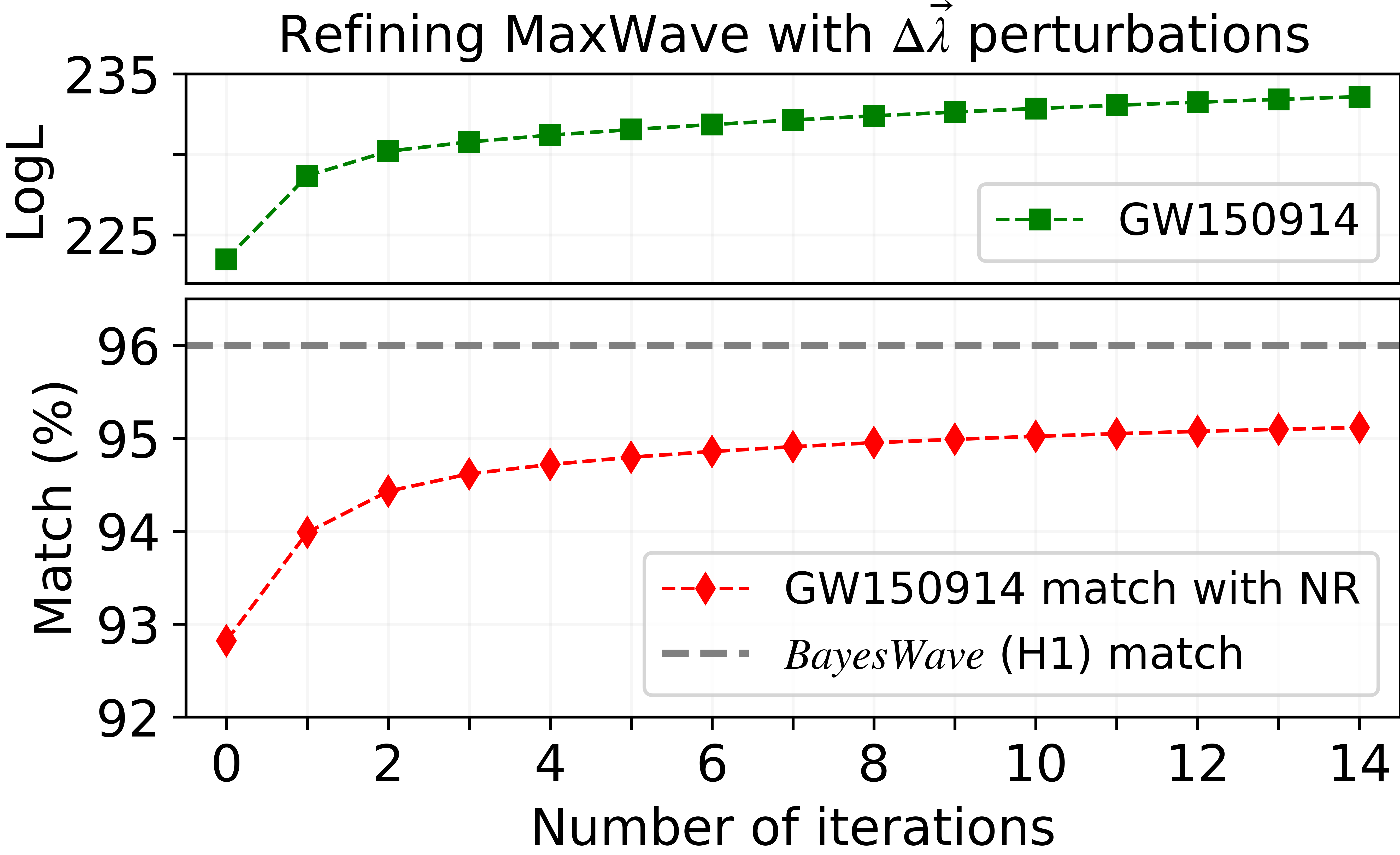}
    \caption{Improvements in $\log{\mathrm{L}}$ (top) and percentage match of MaxWave single-detector recovery with GW150914 NR waveform (bottom) when we iteratively refine the initial approximate maximum likelihood solution with respect to the wavelet parameters $\vec{\lambda}$. Iterative refinements move the MaxWave initial solution off the fixed TF$\tau$ grid to further maximize the $\log{\mathrm{L}}$ (Sec. \ref{subsec:Iterations}). Refined MaxWave improves the single H1 detector match with GW150914 NR waveform from an initial value of 93\% to 95\%. This is comparable to the \textit{BayesWave} RJMCMC single H1 detector match of 96\%.}
    \label{fig:Refined-GW150914}
\end{figure}

The single detector Hanford (H1) match for the MaxWave reconstruction with the publicly released numerical relativity (NR) template for the GW150914 event is 93\%. This is higher than the 91\% single detector match for the old algorithm BayesWave \textit{FastStart}. In Fig. \ref{fig:Refined-GW150914}, we iteratively refine our initial solution off the fixed TF$\tau$ grid to further maximize $\log{\mathrm{L}}$ (as described in Sec. \ref{subsec:Iterations}) and reach a 95\% single detector match with the GW150914 NR waveform. This is comparable to the \textit{BayesWave} RJMCMC single H1 detector match of 96\% and two-detector match of 97\%.

\subsubsection{Operations cost and runtime}\label{subsubsec:Reduction in operations cost}

\hl{We achieve the most significant speedup by eliminating  the cost-intensive step of recomputing wavelet transforms after each subtraction. We analytically subtract overlapping wavelet power for bright pixels by switching to the white wavelet basis and analytic inner products. We further minimize the cost of these analytic subtractions by switching to TF$\tau$ maps, where the power mask is well-localized and the number of bright pixels $N_b \approx 0.002 N_{\text{pixels}}$ for total number of pixels $N_{\text{pixels}} = (N_{\tau} N_f N_t)$. Thus, $ N_b \approx 0.5 N$, which is significantly less as compared to TFQ maps. We also precompute the exponential part of the analytic inner products in Eq.} (\ref{eq:Mij}) \hl{as it is translationally invariant in time and frequency. BayesWave \textit{FastStart}'s cost of recomputing the wavelet transforms to a couple of $e$-folds in frequency $N_{f_Q} \approx \text{min}(20, N_{f_Q})$ after each subtraction is} 
\begin{equation}
    \mathcal{C_\text{N}} = 2 N \log{N} \sum_{N_Q} \text{min}(20, N_{f_Q}).
\end{equation}
\begin{table*}[t]
\centering
\LARGE
\begin{adjustbox}{max width=\textwidth}
\begin{tabular}{|C{0.13\textwidth}|C{0.25\textwidth}|C{0.18\textwidth}|C{0.19\textwidth}|C{0.18\textwidth}|C{0.18\textwidth}|C{0.15\textwidth}|C{0.14\textwidth}|C{0.14\textwidth}|C{0.14\textwidth}|}
\hline
\multirow{2}{*}{\Large\textbf{\boldmath $N_Q=N_\tau$}} &
\multirow{2}{*}{\Large\textbf{Model}} &
\multirow{2}{*}{\Large\shortstack{\textbf{Initial}\\\textbf{Transform}}} &
\multirow{2}{*}{\Large\textbf{Subtractions}} &
\multirow{2}{*}{\Large\shortstack{\textbf{Total Time}\\\textbf{1D}}} &
\multirow{2}{*}{\Large\shortstack{\textbf{Total Time}\\\textbf{3D}}} &
\multirow{2}{*}{\Large\shortstack{\textbf{Refining}\\\textbf{50x}}} &
\multicolumn{3}{|c|}{\Large\shortstack{\textbf{Total time with refinements}}} \\
\cline{8-10}
& & & & & & & \Large\textbf{1D} & \Large\textbf{3D} & \Large\textbf{5D} \\
\hline
\multirow[c]{2}{*}{6} & BW \textit{FastStart} &
0.15 &
1.50 &
1.85 &
\cellcolor{redDark}5.56 &
\dots & \dots & \dots & \dots \\
\cline{2-10}
& MaxWave &
0.06 & 0.01 & 0.28 & 0.83 &
0.26 & 0.54 &
\cellcolor{greenMed}1.62 &
\cellcolor{greenMed}2.70 \\
\hline
\multirow[c]{2}{*}{8} & BW \textit{FastStart} &
0.62 &
3.62 &
\cellcolor{redDark}4.46 &
\cellcolor{redDark}13.38 &
\dots & \dots & \dots & \dots \\
\cline{2-10}
& MaxWave &
0.09 & 0.01 & 0.31 & 0.92 &
0.27 & 0.58 &
\cellcolor{greenMed}1.73 &
\cellcolor{greenMed}2.89 \\
\hline
\end{tabular}
\end{adjustbox}
\caption{\hl{CPU runtime breakdown (in seconds) for BayesWave \textit{FastStart} versus  MaxWave for $N_Q{=}N_\tau{=}6$ and $8$. MaxWave achieves a two order-of-magnitude speedup in wavelet subtractions, reducing \textit{FastStart}’s 1.5–3.6 s per-detector cost to $\leq$ 0.01 s (150–360 $\times$ faster). It also maintains consistent initial wavelet transform times of 0.06–0.09 s ($\approx$10–11 ms per $\tau$ layer) compared to 0.15–0.62 s for \textit{FastStart} as Q increases. The total per-detector runtime are $\approx$1.9 s for \textit{FastStart} and $\approx$0.3 s for MaxWave, including noise spectrum estimation, clustering, approximate maximum likelihood evaluation, and output writing. For a three detector network, \textit{FastStart} exceeds real-time limits ($\approx$5.6 s $>T_{\text{obs}} = 4$ s), while MaxWave --- with up to 50 perturbative refinements per detector --- runs within 1.7 s (three detectors) and 2.9 s (five detectors), maintaining real-time performance.}}
\label{table:Runtimes}
\end{table*}
\hl{For $S_n$ total subtractions, the cost scales by $S_n$. On the other hand, each analytical subtraction in MaxWave involves two cosine calculations [Eq. }(\ref{eq:Mij})] \hl{for every bright pixel that significantly overlaps with the subtracted wavelet. Note that the list of bright pixels above a certain threshold and the overlaps between bright pixels reduce after each subtraction. A conservative estimate of this reduced number of bright pixels after $k$ subtractions is $N_k \approx N_be^{-0.1k}$. As the cost of cosine for $N_k$ bright pixels ranges from $N_k$ to $N_k \log{N_k}$, the upper limit of the $k$th analytical subtraction is about $2 N_k \log{N_k}$. The total cost of $S_k$ analytic subtractions is given by} 
\begin{equation}
    \mathcal{C_\text{A}} = 2 \sum_{k=0}^{S_k} N_k \log{N_k} .
\end{equation}
\hl{Even for the worst case scenario, approximating the total cost of $S_k$ analytic subtractions as $2 {S_k} N_b \log{N_b}$ and assuming that $S_n = S_k$, we get operations of cost ratio}
\begin{equation}
     \frac{\mathcal{C_\text{N}}}{\mathcal{C_\text{A}^{\text{max}}}} = \frac{N \log{N} \sum_{N_Q} \text{min}(20, N_{f_Q})}{N_b \log{N_b}},
\end{equation}
\hl{which is more than 200 times (320 times) faster for $N_Q = 6$ ($N_Q = 8$). Note that the operations cost for analytical subtractions does not directly scale with $N_{\tau}$. 

Downsampling and heterodyning the initial wavelet transform further reduces the runtime. The TFQ map in the old algorithm is logarithmically sampled in frequency to produce a minimum frequency overlap of 0.950 [calculation similar to Eq.} (\ref{eq:overlaps})]. \hl{Thus $\Delta \ln{f} \propto 1/Q$ and $N_{f_Q} \propto Q$. The operations cost of the old algorithm's initial wavelet transform in given as} 
\begin{equation}
    C_{\text{wt}}^{(Q)} = 2 N \log{N} \sum_{N_Q} N_{f_Q}. 
\end{equation}
\hl{In contrast, the cost of MaxWave's downsampled, heterodyned initial wavelet transform is} 
\begin{equation}
    C_{\text{wt}}^{(\tau)} = \sum_{N_{\tau}} 2 N_f N_t \log{N_t} = 2 N_{\text{grid}} \sum_{N_{\tau}} \log{N_t}.
\end{equation}
\hl{For $N_Q{=}6$ and $Q\in [1, 32]$, $\sum_{N_Q} N_{f_Q}{=}236$. Comparing this with the $N_\tau = 6$ case, we get $C_{\text{wt}}^{(Q)}/ C_{\text{wt}}^{(\tau)} \approx 1.5 $. This speedup is even more significant if we further increase $N_Q$ as $\sum_{N_Q} N_{f_Q}$ grows significantly when we include even larger Q values. 

We also compare the CPU runtimes of BayesWave \textit{FastStart} and MaxWave on 4 seconds of LIGO data sampled at 4096 Hz, and downsampled by 2 to give $N=8192$. All CPU times are measured on a single core Intel i7-7820HQ (2.9 GHz) processor. Table} \ref{table:Runtimes} \hl{shows that MaxWave achieves substantially lower runtimes, maintaining real-time performance across all configurations. The largest gain appears in wavelet subtractions; \textit{FastStart} requires about 1.5 s per detector for $N_Q{=}6$ and about 3.6 s for $N_Q{=}8$, while MaxWave completes analytical subtractions in $\leq$ 0.01 s, yielding a 150–360x speedup. For the initial wavelet transform, \textit{FastStart} takes about 0.15–0.62 s as $N_Q$ increases, driven by the dramatically growing $N_{f_Q}$ for larger Q values, while MaxWave maintains a nearly constant speed at 0.06–0.09 s for increasing $N_\tau$ (about 10–11 ms per $\tau$ layer). Since runtime scales linearly with $T_{\text{obs}}$, real-time analysis requires the total runtime to be less than $T_{\text{obs}}{=}4$ s. The total per-detector runtime is about 1.9 s for BayesWave \textit{FastStart} and about 0.3 s for MaxWave. This includes noise spectrum estimation ($\approx$0.2 s), clustering ($\approx$0.0002 s), approximate maximum likelihood evaluation ($\approx$0.004 s), and output writing ($\approx$0.0002 s). It does not include the I/O overhead from reading the noise spectrum and strain data, which can be optimized within LIGO pipelines. For a three detector network, \textit{FastStart} exceeds real-time limits ($\approx$5.6 s $> T_{\text{obs}})$, while MaxWave --- with up to 50 perturbative refinements per detector --- runs within 1.7 s for three detectors and 2.9 s for five detectors and remains capable of real-time analysis. 

Extending MaxWave to a coherent signal model will introduce one additional detector formed by aligning the individual detectors through z-statistic based time and phase shifts and amplitude scalings. Subtractions, clustering, and the approximate maximum likelihood evaluation on this additional detector will add at most 0.2 s to the total runtime, keeping the future MaxWave signal model well within real-time analysis requirements.}

\begin{figure*}[t]
    \centering
    \includegraphics[width=17cm]{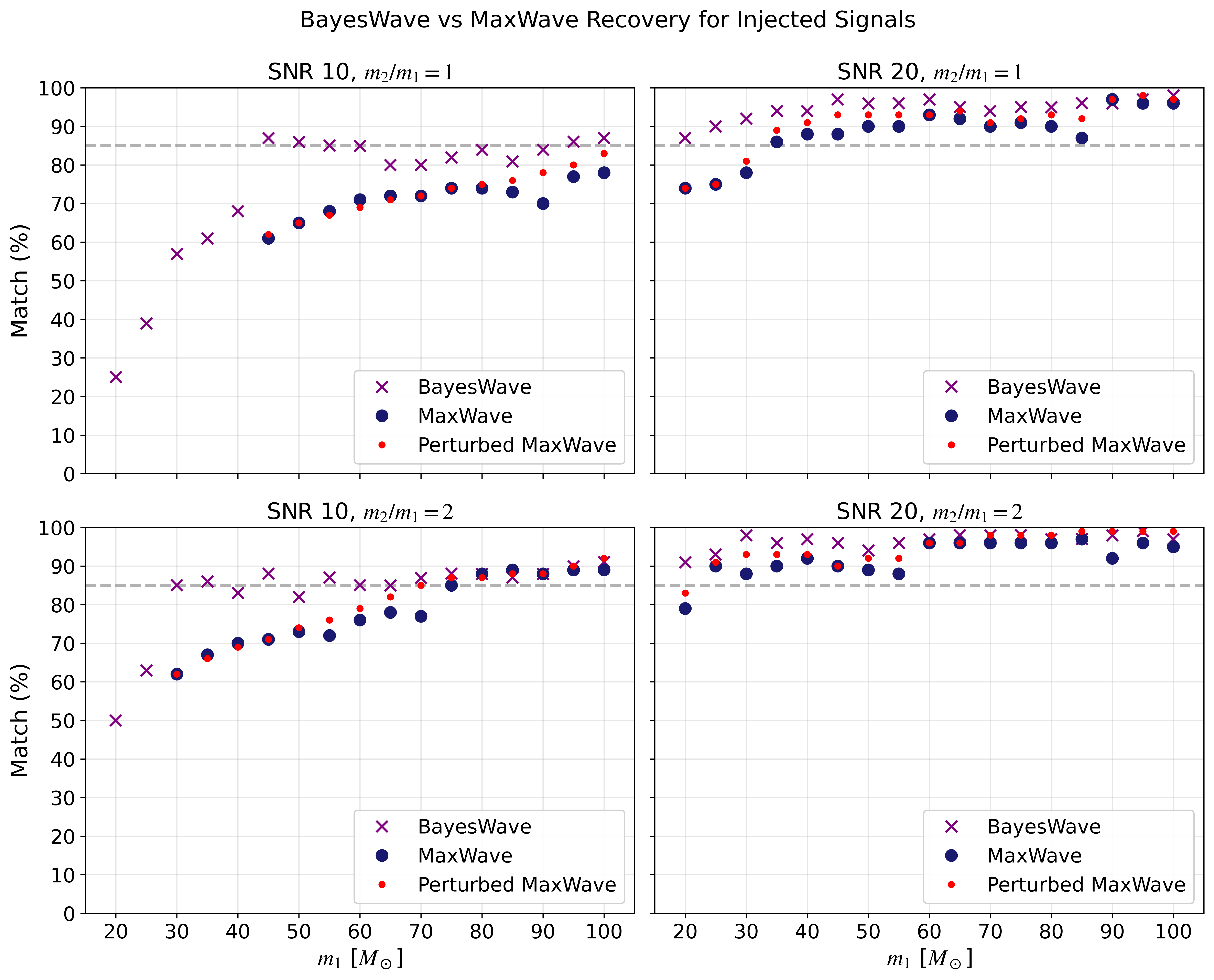}
    \caption{Comparing matches of \textit{BayesWave} (cross) and MaxWave (circles) recoveries for SNR 10 (left) and SNR 20 (right) IMRPhenomD binary black hole injections with $m1 \in (20, 100)M_\odot$ and mass ratios $m_2/m_1 {=} 1$ (top) and $2$ (bottom) in real LIGO noise at GPS trigger time 1268903507 \cite{GWOSC_O1_O3, GWTC-3}. The refined MaxWave solution is computed by iteratively maximizing the initial $\log{\mathrm{L}}$ as described in Sec. \ref{subsec:Iterations}. Maximizing $\log{\mathrm{L}}$ can lead us to fit more noise and does not always improve the matches of the recovered waveforms with the corresponding injected templates. This is especially true for low SNR signals. \hl{Compared to the robust but computationally intensive \textit{BayesWave}, MaxWave may trade reconstruction accuracy to provide a real-time, low-latency, approximate maximum likelihood solution. However, as we increase SNR and individual binary masses, recovered waveforms matches for refined MaxWave (red circles) converge towards the \textit{BayesWave} recovery.} The dips in the match with increasing $m_1$ show us that both \textit{BayesWave} and MaxWave are sensitive to how the injected waveform aligns with specific features of the noise segment. The fluctuations disappear when we average over MaxWave matches for signals injected in 16 different real LIGO noise realizations in Fig. \ref{fig:Averaged_Injections}.}
    \label{fig:Injections}
\end{figure*}

\subsection[\textit{BayesWave}]{MaxWave versus \textit{BayesWave} RJMCMC}\label{subsec:MW vs BW} 

\subsubsection{\textup{Injected signal recovery}}

\begin{figure*}[t]
     \centering
    \includegraphics[width=17cm]{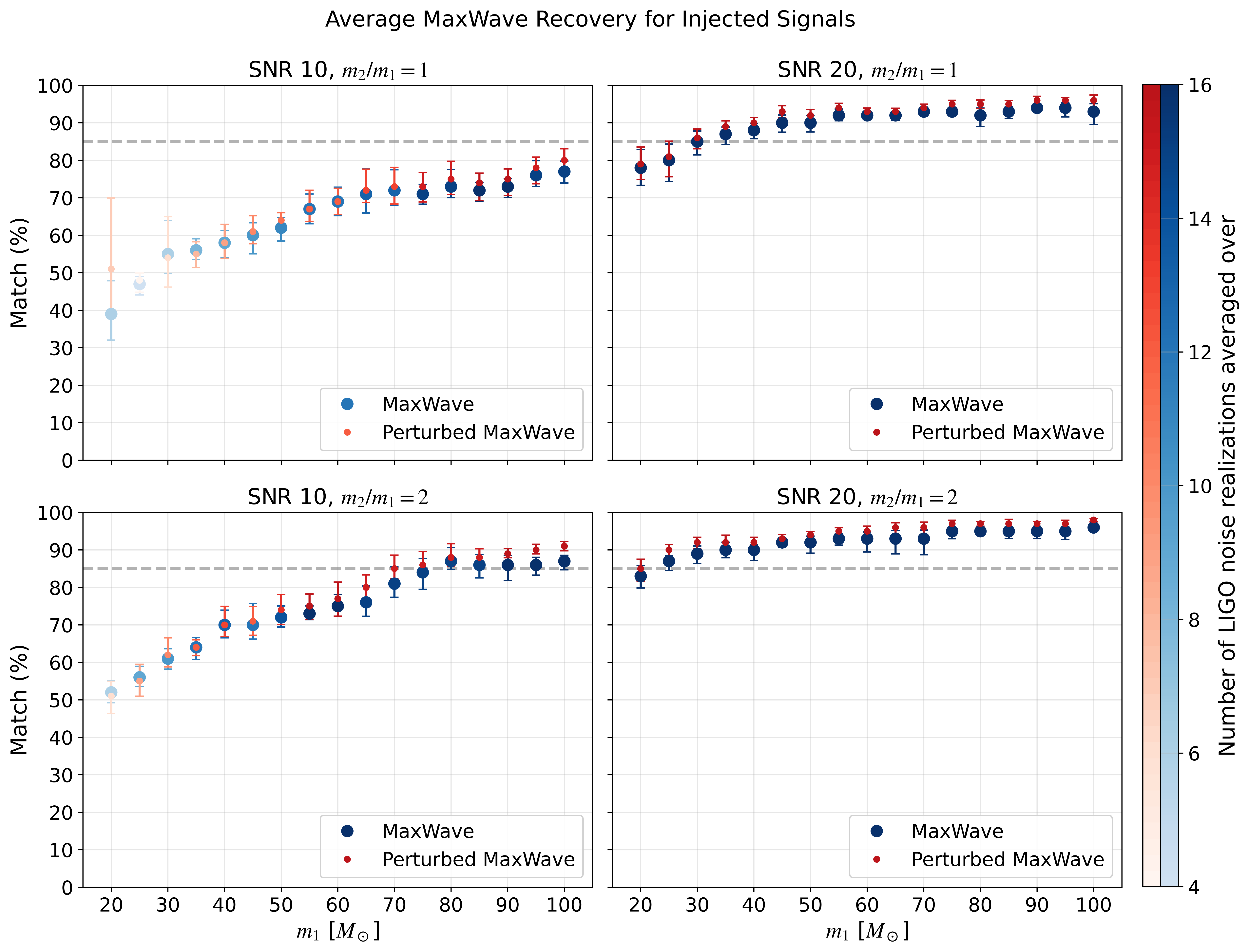}
    \caption{Averaged matches of MaxWave initial and refined recoveries for SNR 10 (left) and SNR 20 (right) IMRPhenomD binary black hole injections with $m1 \in (20, 100)M_\odot$ and mass ratios $m_2/m_1 {=} 1$ (top) and $2$ (bottom). The signals are injected in 16 different real LIGO noise realizations with GPS trigger times between 1265235987-91, 1266214778-82, 1267724178-82, and 1268903502-07 \cite{GWOSC_O1_O3, GWTC-3}. For low SNR, MaxWave does not always recover a wavelet. The colorbar indicates an average over the number of times --- out of the injections in 16 different LIGO noise realizations --- that MaxWave found a wavelet. The blue (red) data points indicate averaged match for MaxWave (refined MaxWave) recoveries. The corresponding error bars give the asymmetrical positive and negative standard deviation around the averaged match. For refined MaxWave on these 16 injections, we observe a $1.6^{+1.8}_{-1.7}$\% improvement in the match for SNR 10 injections and a $2.2^{+1.1}_{-1.2}$\% improvement in match for SNR 20 injections. \hl{Similar to \textit{BayesWave}}\cite{Pannarale_2019, Neil-BWstart-2021}, \hl{MaxWave performs better for higher-mass binaries as they occupy smaller time–frequency volumes and localize power within a few higher-SNR wavelets.}}
    \label{fig:Averaged_Injections}
\end{figure*}

We inject SNR 10 and 20 IMRPhenomD binary black hole signals with $m1 \in (20, 100)M_\odot$ and mass ratios $m_2/m_1 {=} 1$ and $2$ in real LIGO noise at GPS trigger time 1268903507 \cite{GWOSC_O1_O3, GWTC-3}. In Fig. \ref{fig:Injections}, we compare the match between the injected and recovered waveforms for \textit{BayesWave} (crosses) and MaxWave initial and refined solutions (circles). The left (right) column plots matches for SNR 10 (SNR 20) injections. The first (second) row plots matches for mass ratio $m_2/m_1 {=} 1$ ($m_2/m_1 {=} 2$). The refined MaxWave solution is computed by iteratively maximizing the initial $\log{\mathrm{L}}$ as described in Sec. \ref{subsec:Iterations}. Maximizing $\log{\mathrm{L}}$ can lead us to fit more noise and does not always improve the matches of the recovered waveforms with the corresponding injected templates. This is especially true for low SNR signals. 

\hl{We expect the match between the injected and recovered waveforms for MaxWave (circles) to be lower than those for \textit{BayesWave} (crosses). While \textit{BayesWave} reconstructs signals through a computationally intensive RJMCMC with trans-dimensional sampling and requires hours to run, MaxWave provides a real-time, low-latency, approximate maximum likelihood solution. Consequently, MaxWave may trade reconstruction accuracy for substantial computational efficiency. However, as we increase SNR and individual binary masses, recovered waveforms matches for refined MaxWave (red circles) converge towards the \textit{BayesWave} recovery.}

The dips in the match with increasing $m_1$ in Fig. \ref{fig:Injections} are incidental. These variations show us that both \textit{BayesWave} and MaxWave are sensitive to how the injected waveform aligns with specific features of the noise segment. Small changes in waveform morphology can interact differently with the noise, leading to fluctuations in the recovered match. 

The fluctuations disappear when we average over MaxWave matches for signals injected in 16 different real LIGO noise realizations with GPS trigger times between 1265235987-91, 1266214778-82, 1267724178-82, and 1268903502-07 \cite{GWOSC_O1_O3, GWTC-3} (Fig. \ref{fig:Averaged_Injections}). For low SNR, MaxWave does not always recover a wavelet. The colorbar indicates an average over the number of times --- out of the injections in 16 different LIGO noise realizations --- that MaxWave found a wavelet. The blue (red) data points indicate averaged match for MaxWave (refined MaxWave) recoveries. The corresponding error bars give the asymmetrical positive and negative standard deviation around the averaged match. For refined MaxWave on these 16 injections, we observe a $1.6^{+1.8}_{-1.7}$\% improvement in the match for SNR 10 injections and a $2.2^{+1.1}_{-1.2}$\% improvement in the match for SNR 20 injections. 

\hl{At fixed SNR, both \textit{BayesWave} (crosses in Fig.} \ref{fig:Injections}) \hl{and MaxWave (Fig.} \ref{fig:Averaged_Injections}) \hl{perform better for higher-mass binaries, which occupy smaller time–frequency volumes, localize power within a few higher-SNR wavelets and are more efficiently reconstructed. On the other hand, lower-mass binaries span many cycles across a wide frequency range, distribute SNR over a large time–frequency volume, increase the number of low-SNR wavelets needed to reconstruct the signal, and reduce reconstruction fidelity. This behavior is consistent with the known performance of \textit{BayesWave}} \cite{Pannarale_2019, Neil-BWstart-2021}. \hl{Any improvement in recovery with increasing total mass is limited by the detector noise spectrum, as extremely massive systems below the sensitive frequency band are less recoverable.}

\subsection{Time and frequency domain glitch reconstructions}\label{subsec:TF reconstructions of Glitches}

\begin{figure*}[t]
    \raggedright 
    \includegraphics[width=3cm]{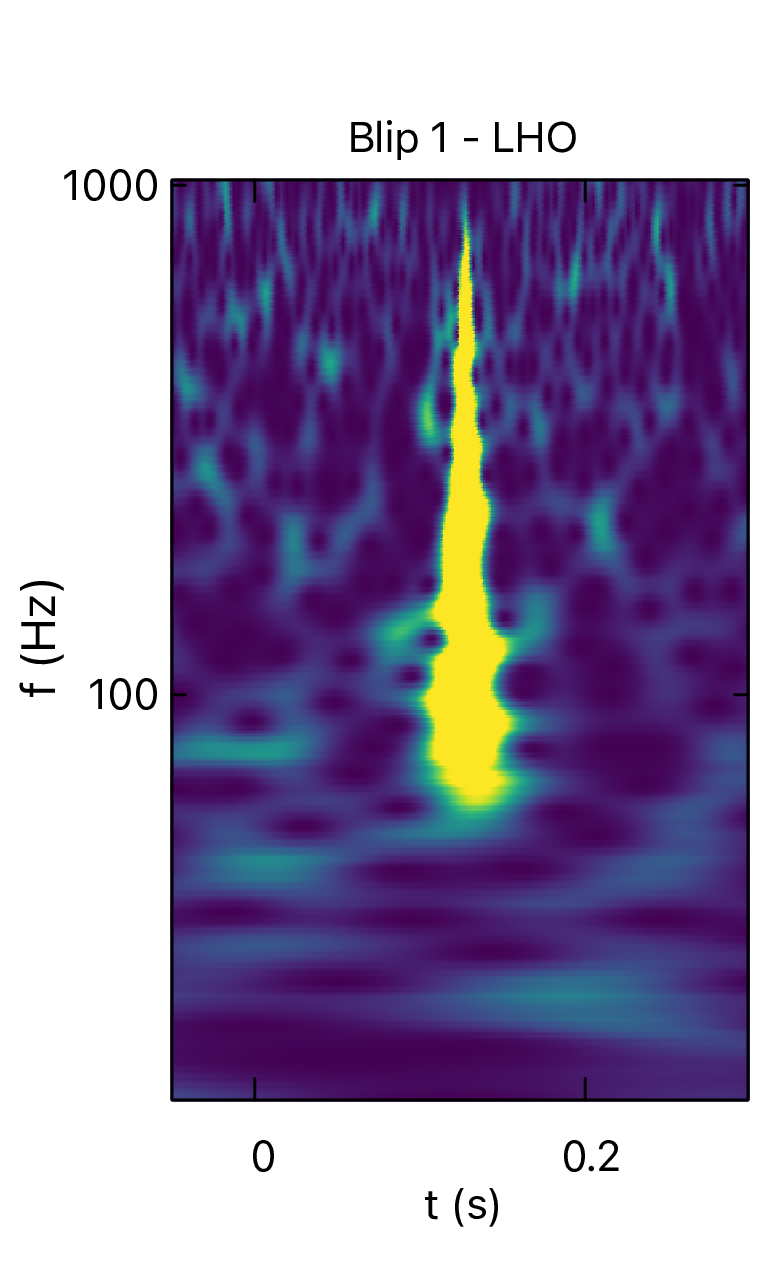}
    \raisebox{2.25cm}{
    \parbox[c]{4.2cm}{
     \includegraphics[width=4.15cm, height=2.4cm]{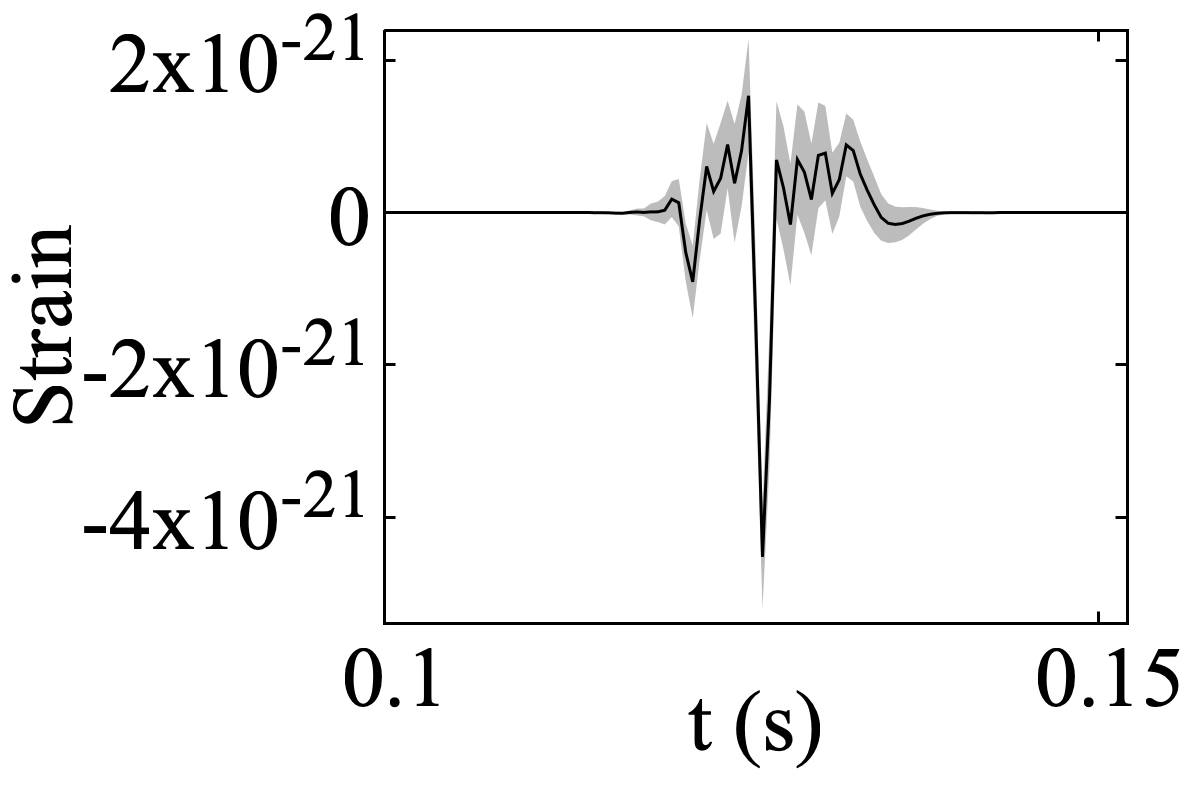}
     \includegraphics[width=4.2cm, height=2.15cm]{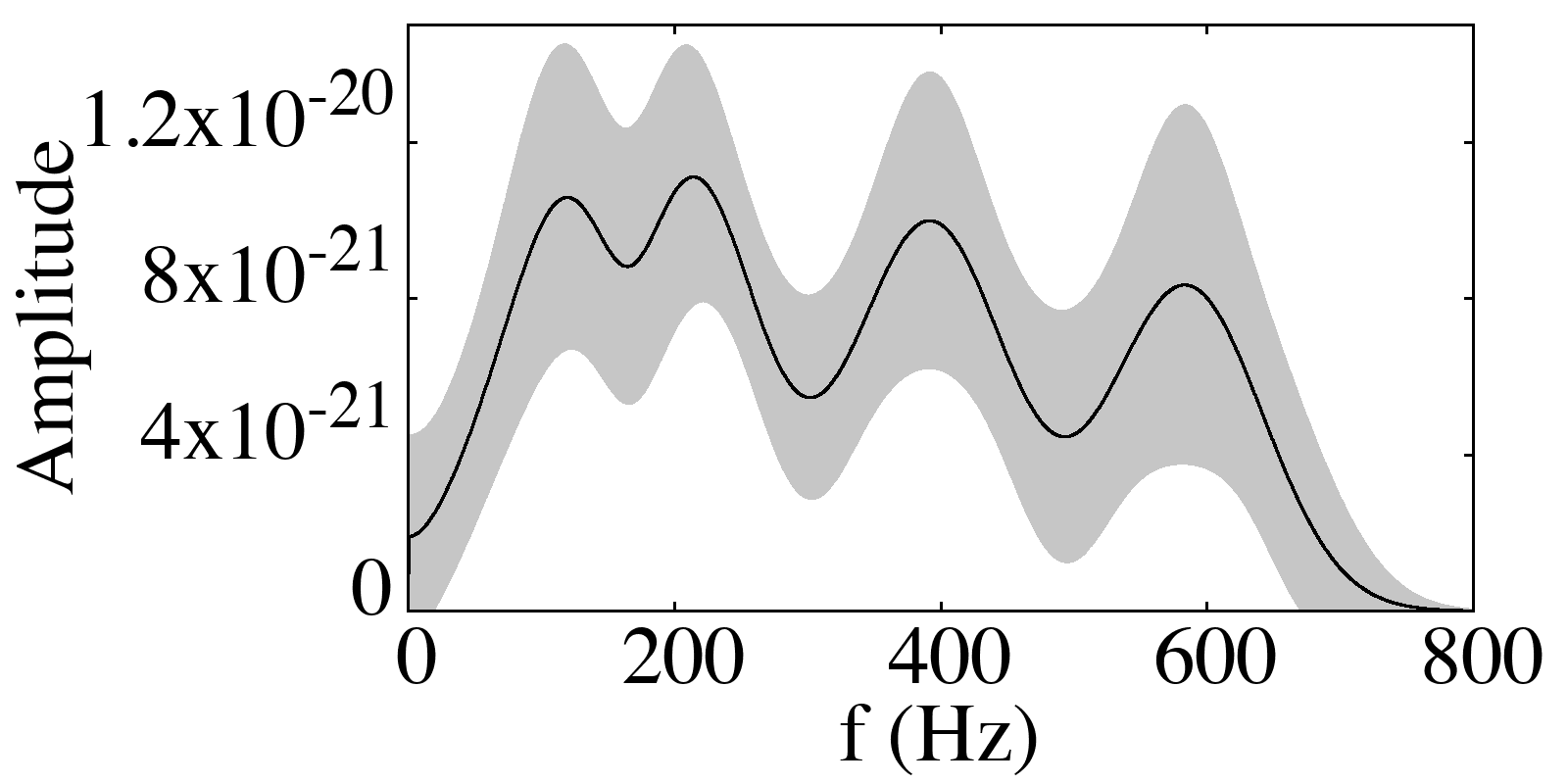}
     }}
     \raisebox{1cm}{
     \includegraphics[width=4.5cm]{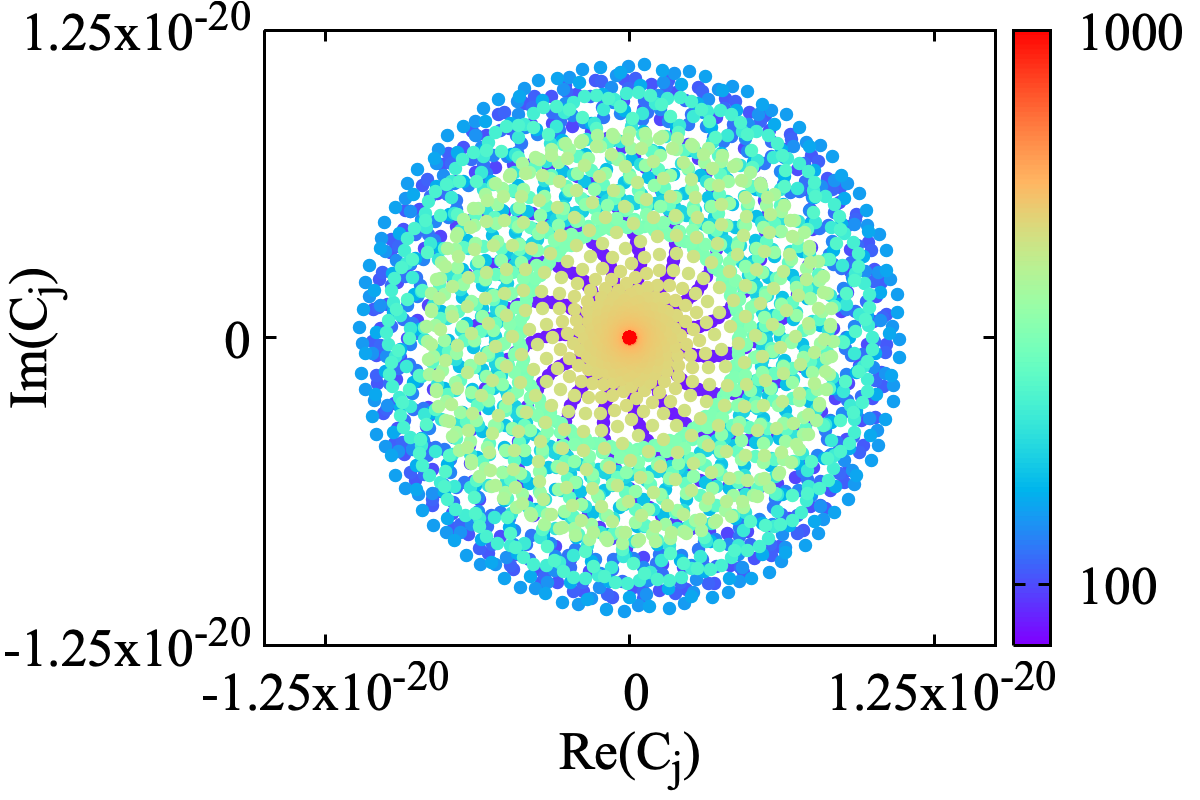}
     }
     \includegraphics[width=5.5cm]{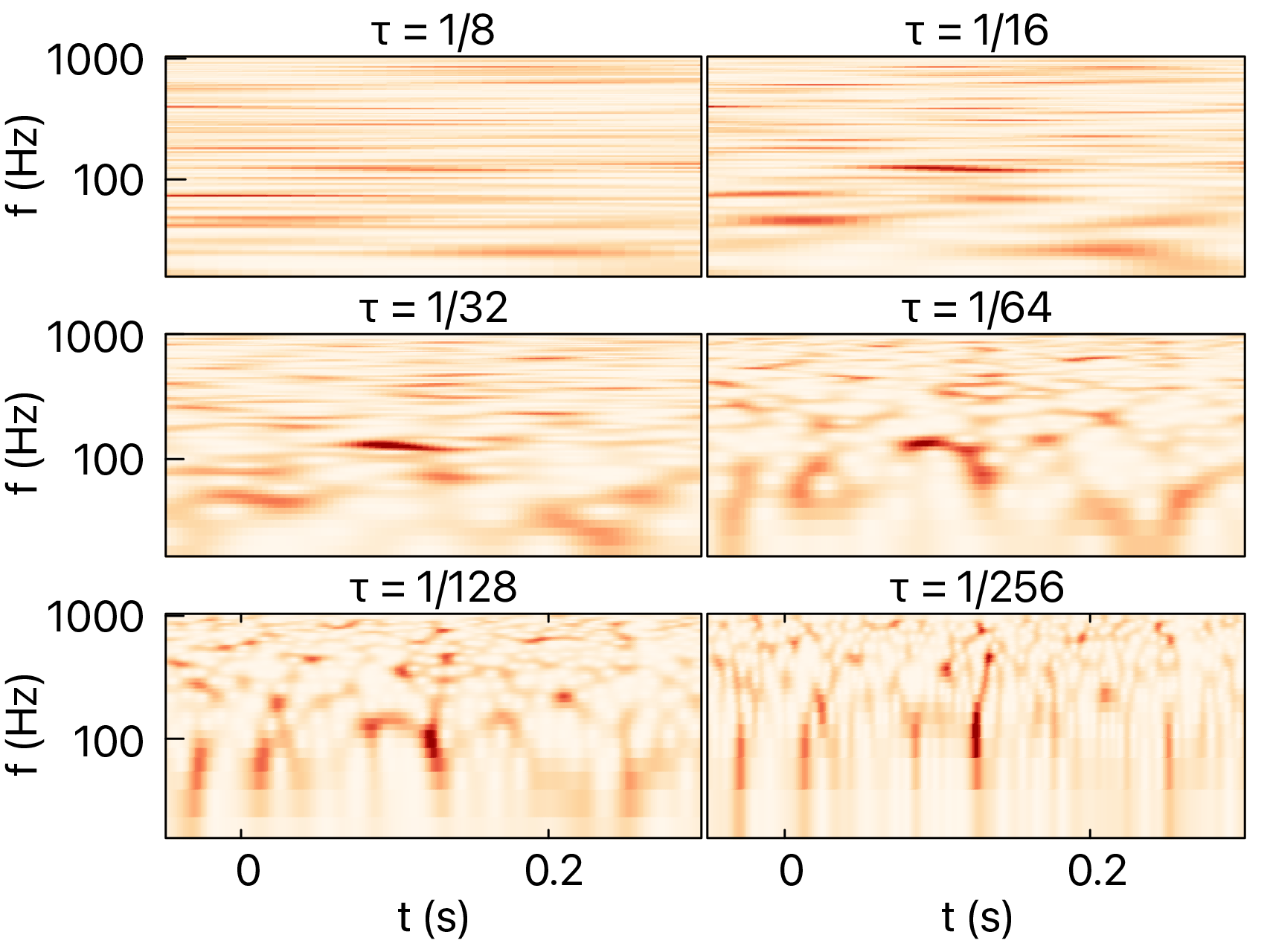}
     
    \includegraphics[width=3cm]{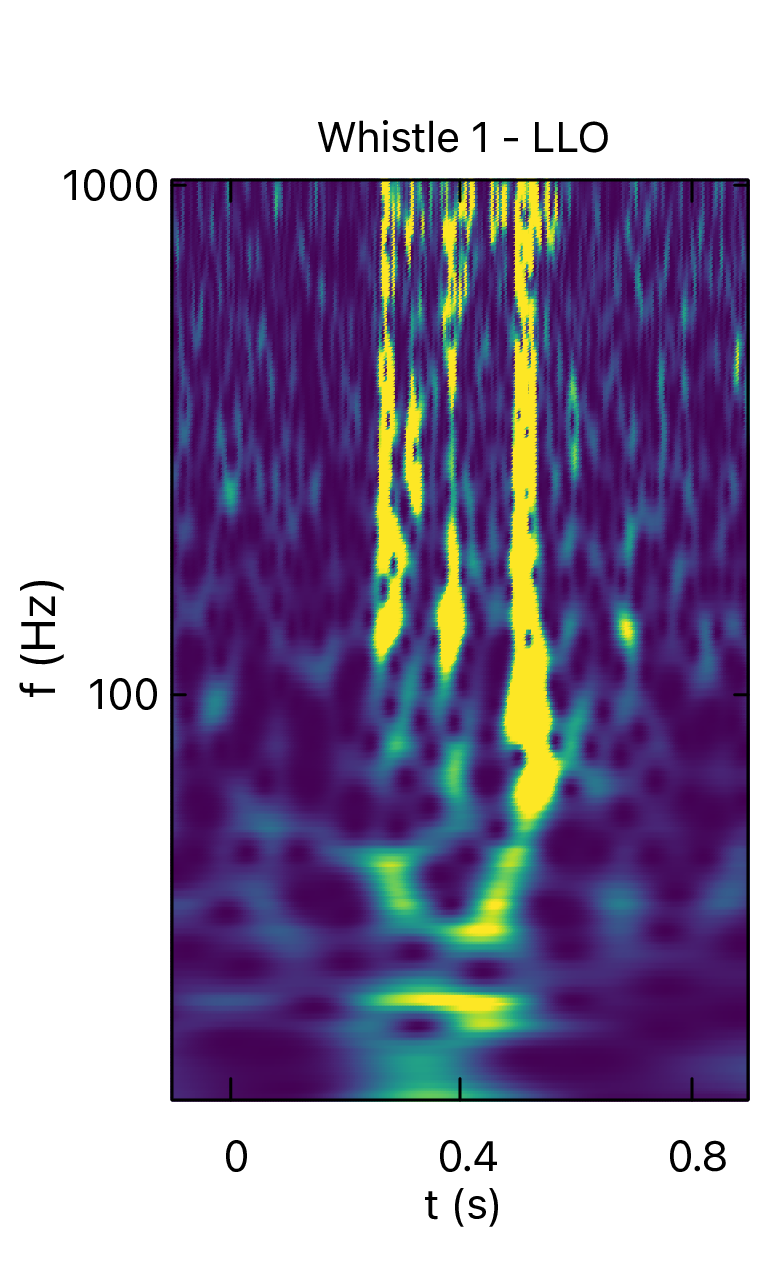}
    \raisebox{2.25cm}{
    \parbox[c]{4.2cm}{
     \includegraphics[width=4.15cm, height=2.5cm]{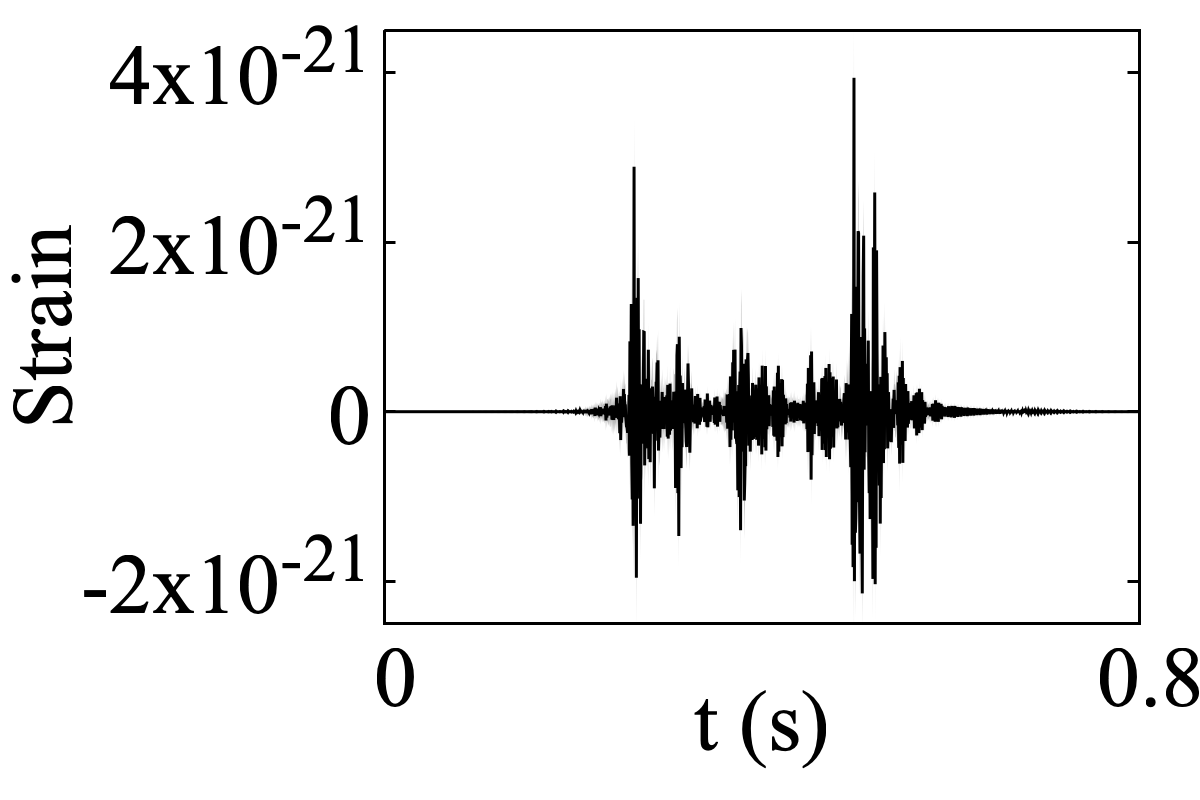}
     \includegraphics[width=4.2cm, height=2.15cm]{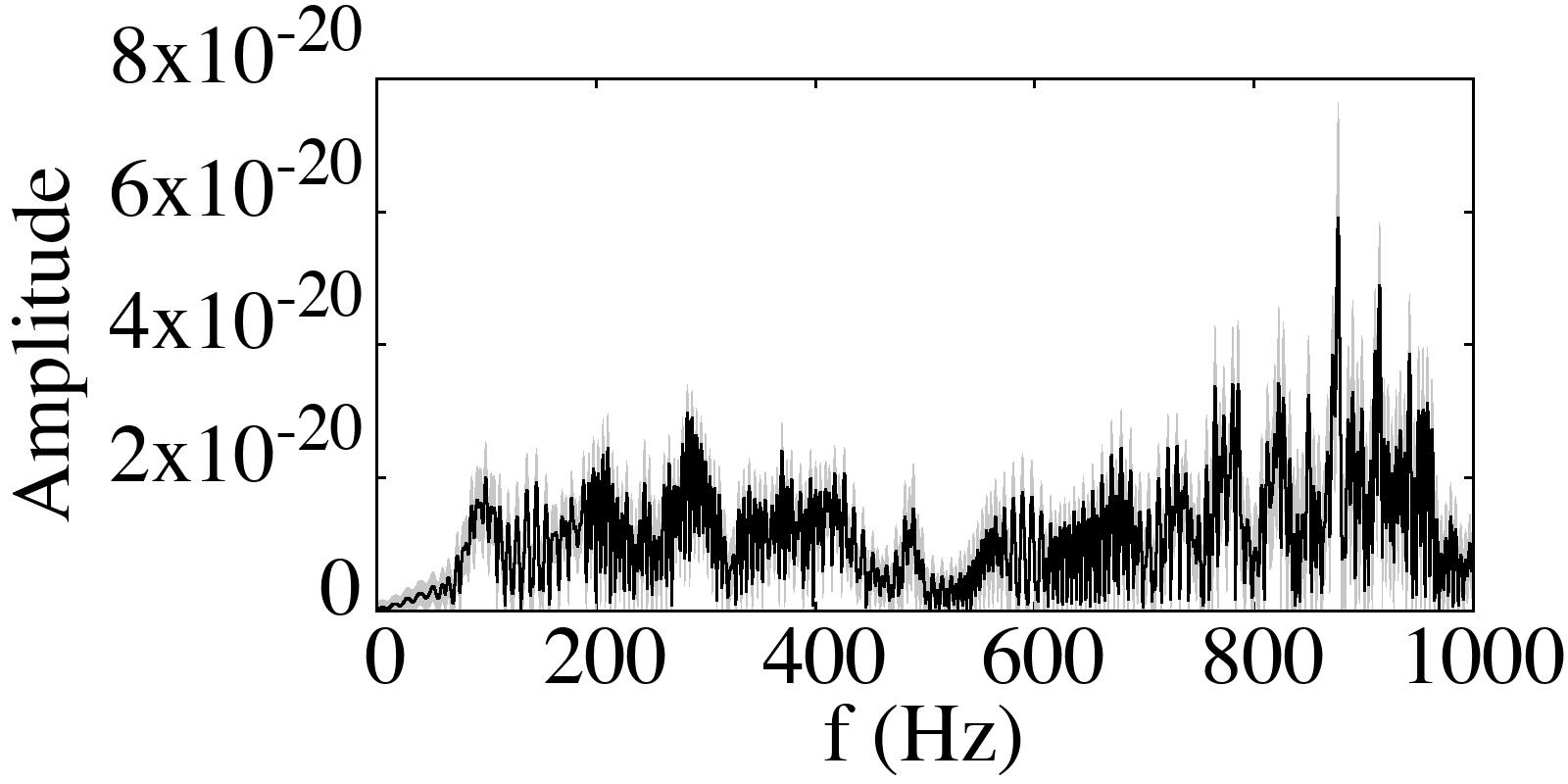}
     }}
     \raisebox{1cm}{
     \includegraphics[width=4.5cm]{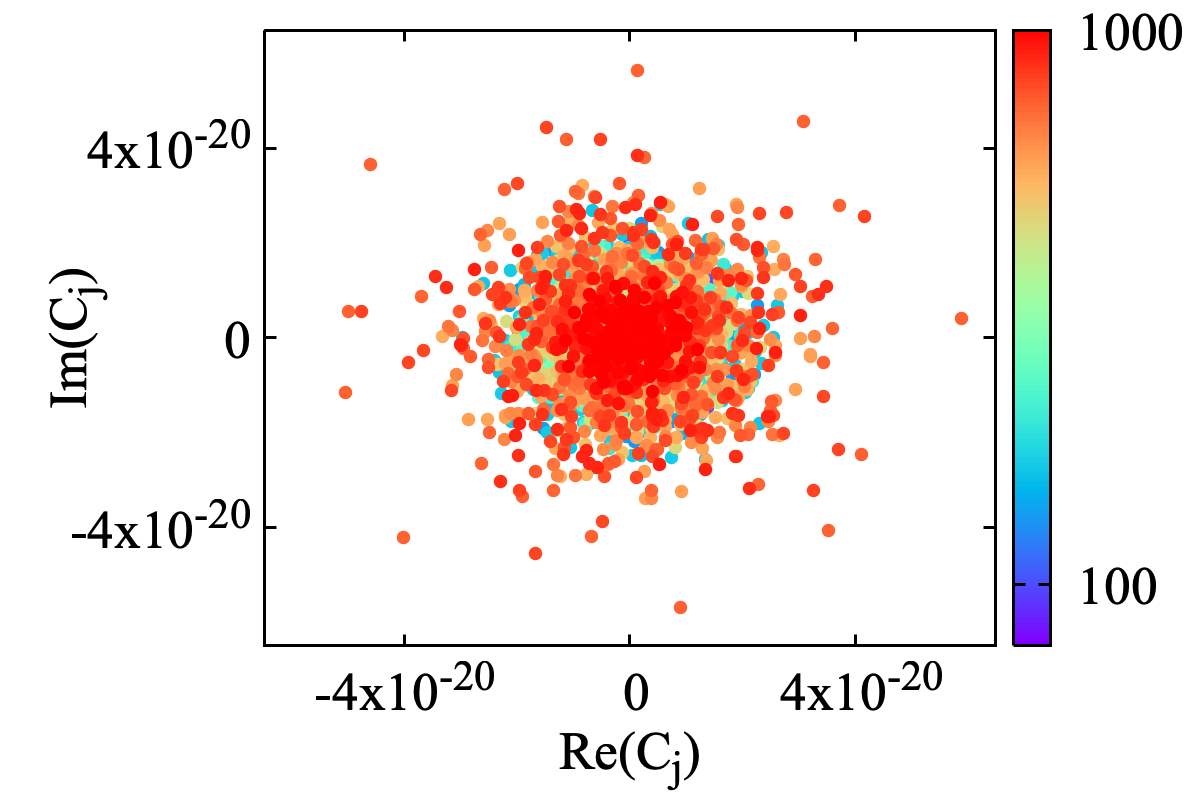}
     }
     \includegraphics[width=5.5cm]{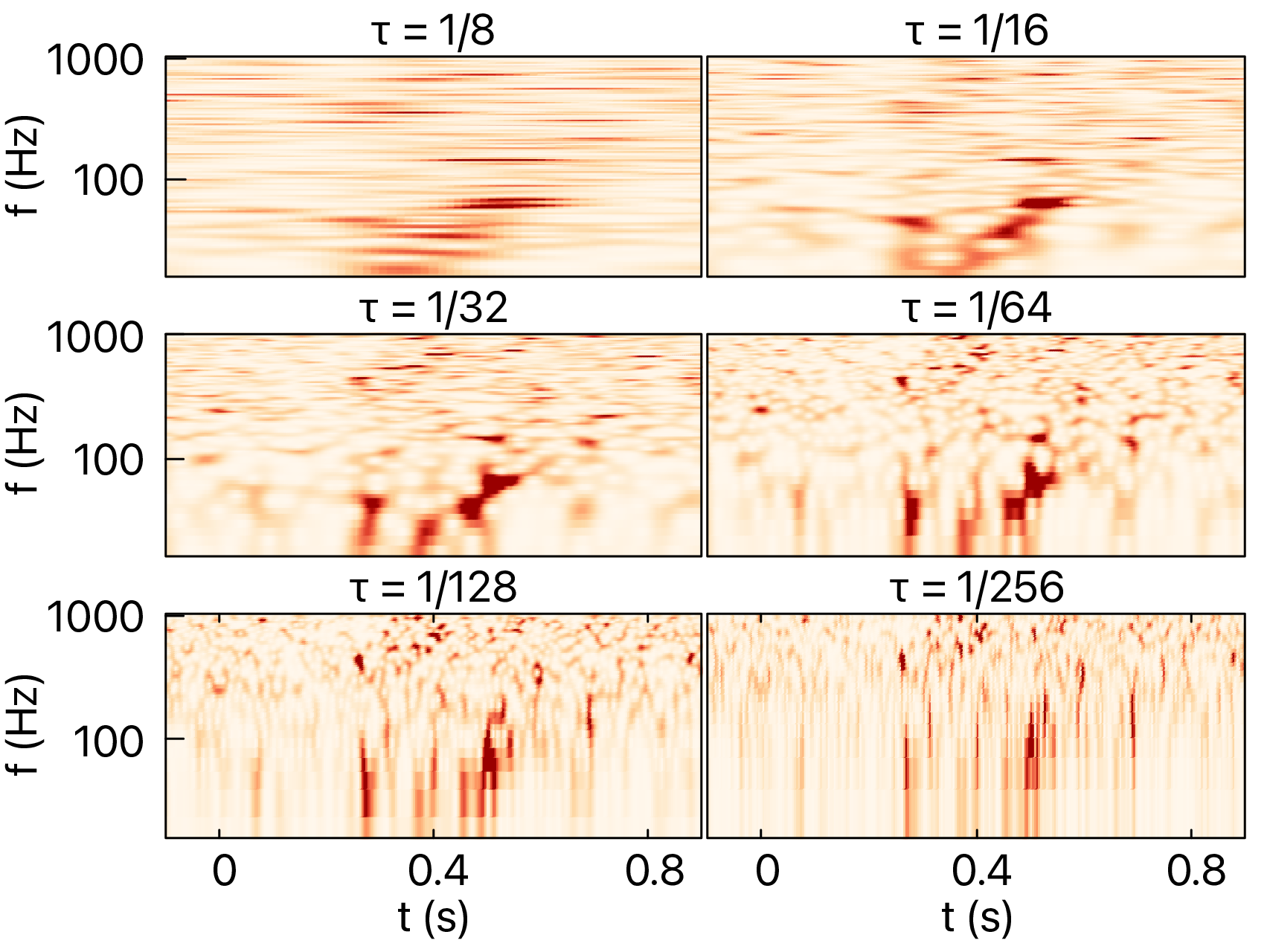}
    
    \includegraphics[width=3cm]{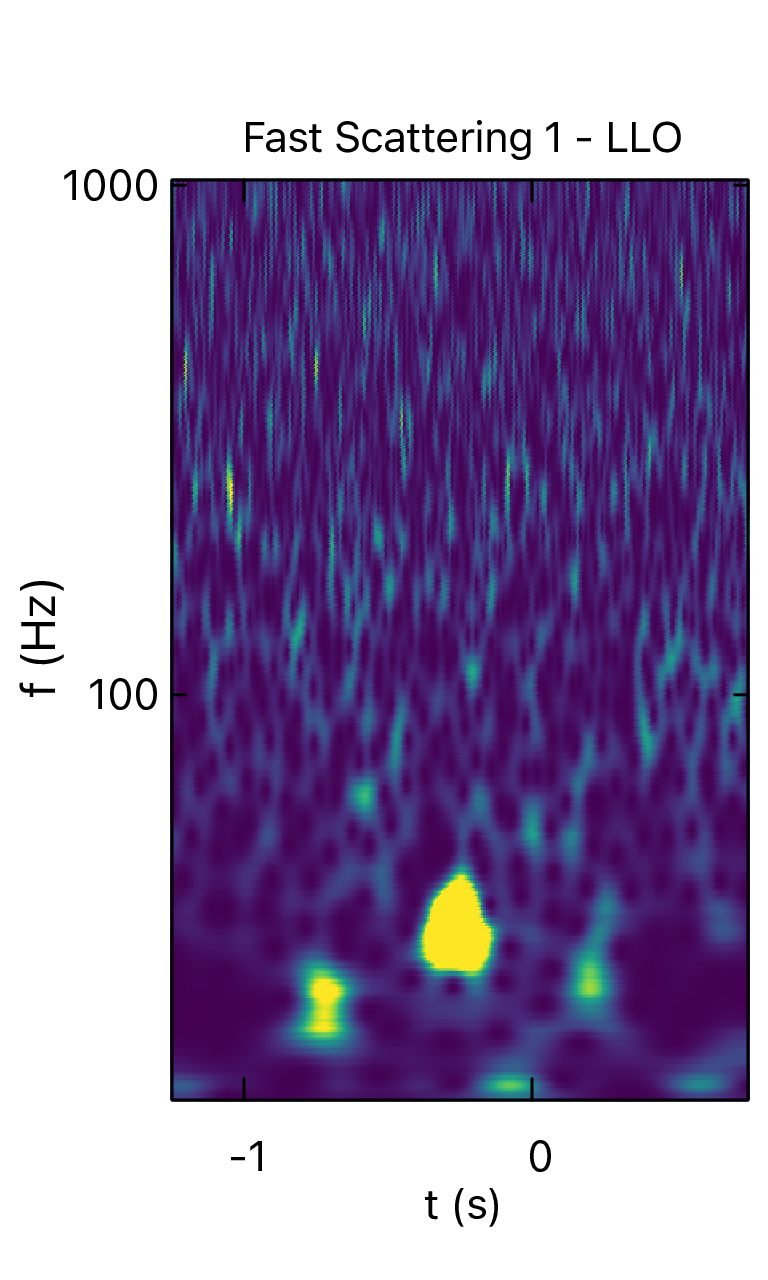}
    \raisebox{2.25cm}{
    \parbox[c]{4.2cm}{
     \includegraphics[width=4cm, height=2.4cm]{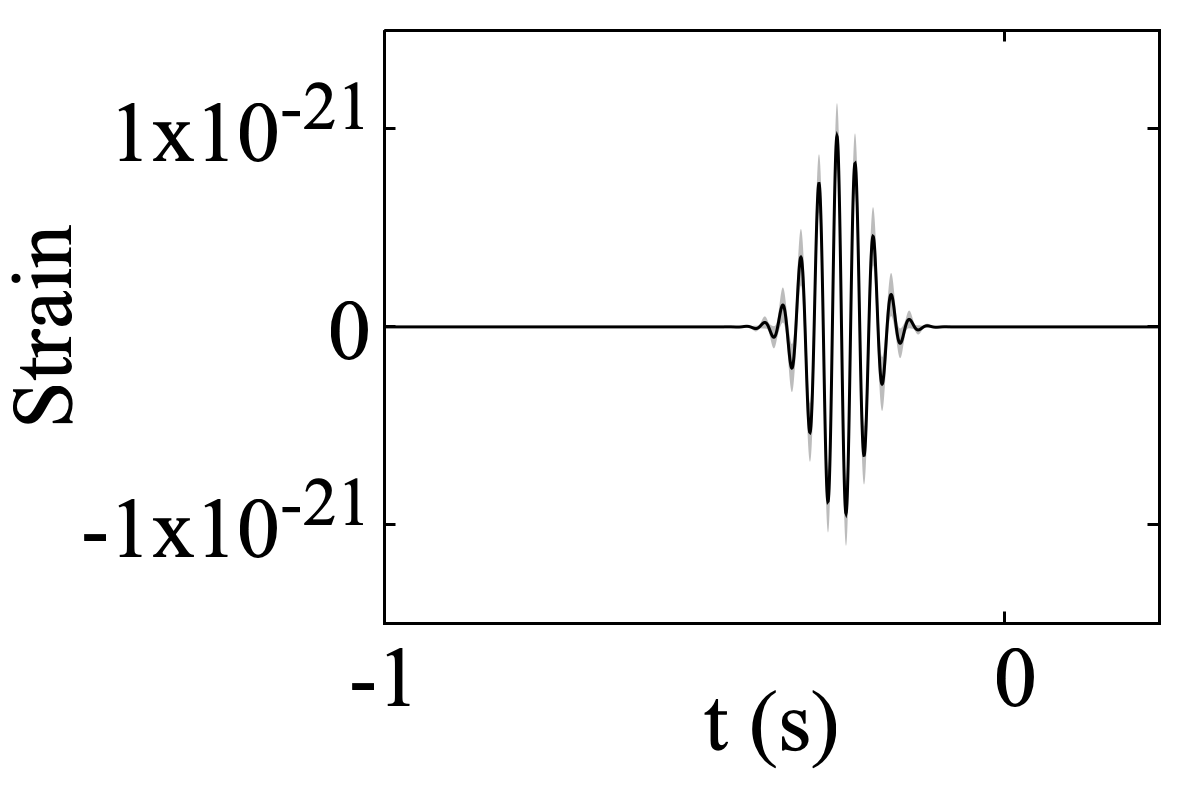}
     \includegraphics[width=4.2cm, height=2.15cm]{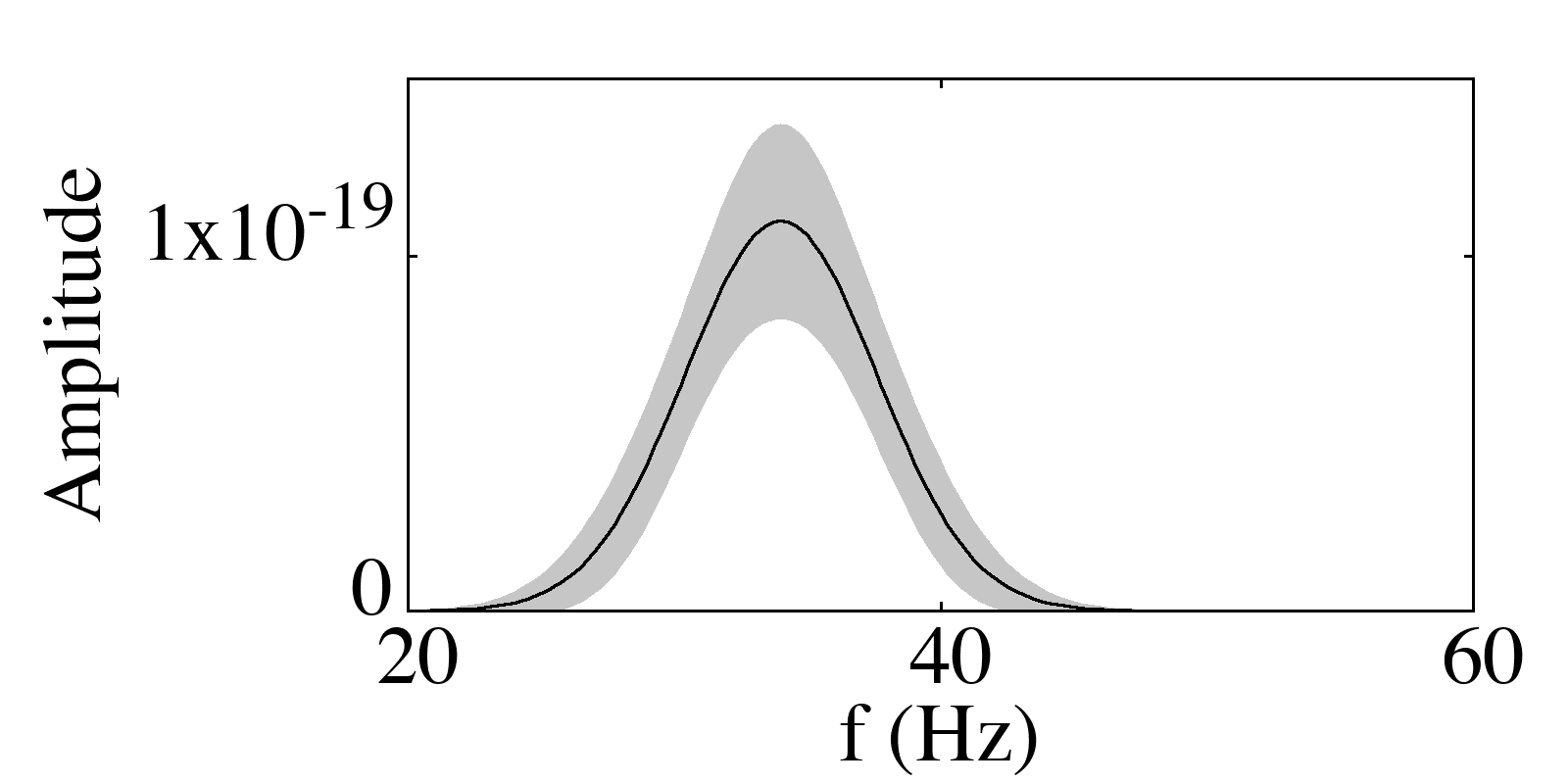}
     }}
     \raisebox{1cm}{
     \includegraphics[width=4.5cm, height=3.2cm]{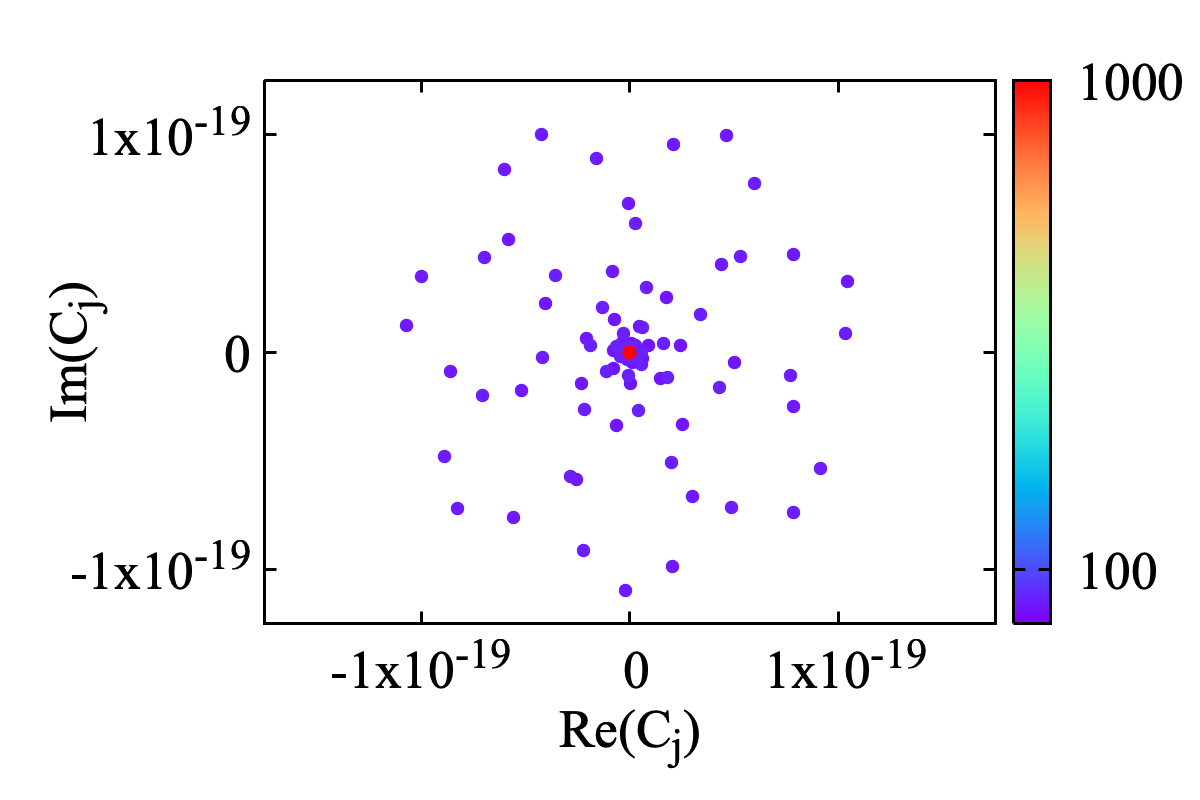}
     }
     \includegraphics[width=5.5cm]{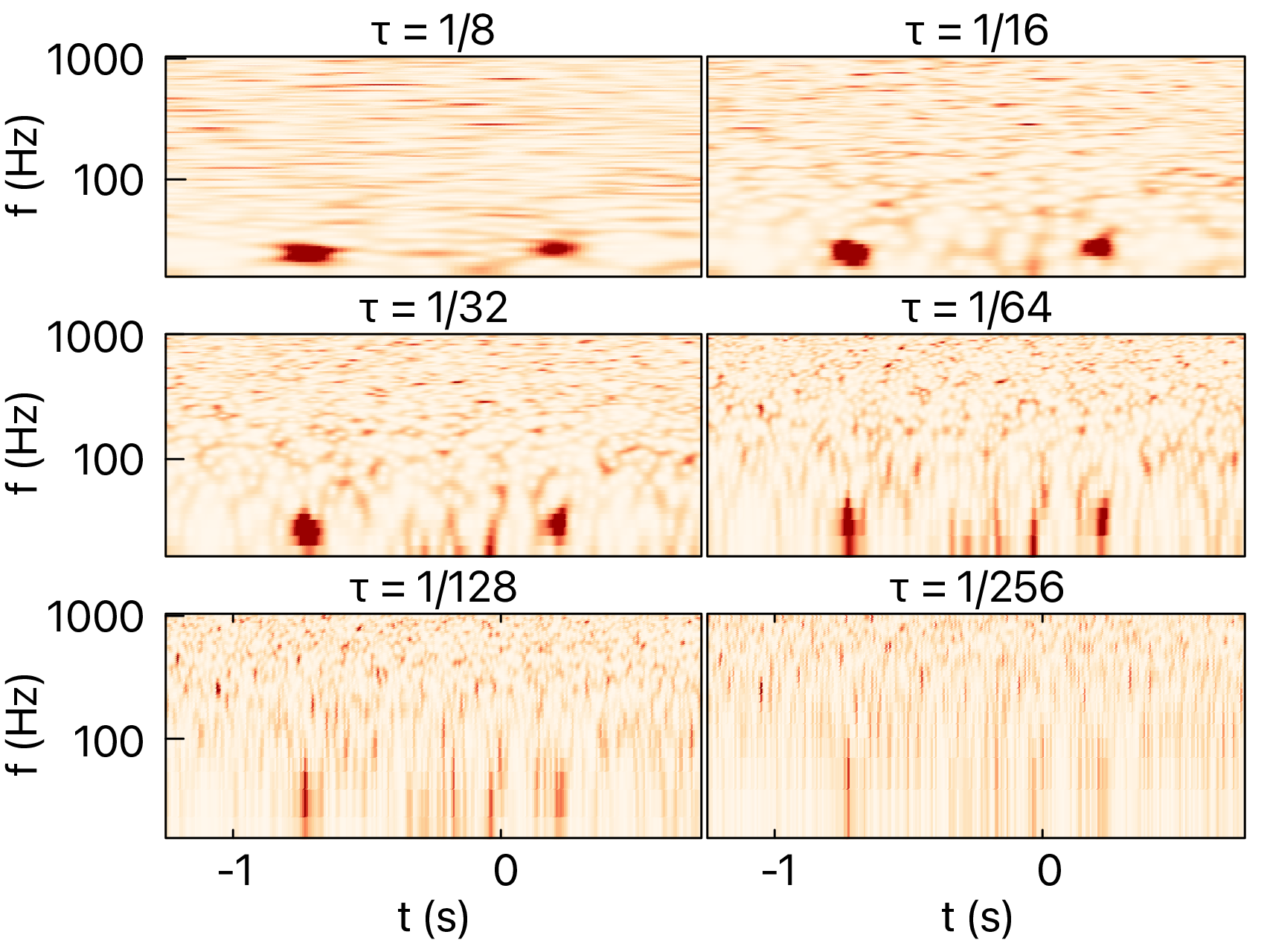}
    \caption{Spectrograms/ Qscans (column 1) \cite{Sophie-2022}, time and frequency domain nonwhitened reconstructions (columns 2 and 3) and residuals (column 4) of various types of glitches. Our reconstructions depend on the number of wavelets identified and the chosen time-frequency resolution. They should only be compared with noise models subject to the detector's frequency sensitivities. MaxWave struggles with subtracting glitches at low SNRs, which can be confused with Gaussian noise outliers.}
    \label{fig:Glitches}
\end{figure*}

We render the time and frequency domain nonwhitened reconstructions of different types of glitches in columns 2 and 3 of Fig. \ref{fig:Glitches}. We also plot the residual after glitch subtraction in column 4. The reconstructions should only be compared with noise models subject to the detector's frequency sensitivities. Note that MaxWave struggles with subtracting glitches at low SNRs, which can be confused with Gaussian noise outliers.

The extra information about a glitch's time and frequency domain features can greatly aid Gravity Spy ML algorithms \cite{Gravity_Spy_2017}, which are currently only using spectrograms/Qscans (column 1 of Fig. \ref{fig:Glitches}) to understand instrument noise. \hl{The data whitening process and Gaussian background in Qscans can alter the apparent morphology of glitches and affect how ML models interpret them. Although our nonwhitened reconstructions are also altered by the number of wavelets identified and the chosen time-frequency resolution, they isolate the non-Gaussian component of any glitch and enrich its morphological information. When incorporated into training data alongside the Qscans, these reconstructions can enhance glitch classification and improve the performance of spectrogram-based ML models.}

\section{CONCLUSION}\label{sec:Conclusion}

\hl{In this paper, we demonstrate that MaxWave, our rapid, approximate maximum likelihood wavelet reconstruction algorithm, achieves real-time reconstructions of non-Gaussian artifacts across multidetector networks. The most significant speedup --- over two orders of magnitude --- arises from adopting a modified wavelet basis with analytical inner-product subtractions, eliminating BayesWave \textit{FastStart}’s cost-intensive, iterative recomputation of wavelet transforms} \cite{Neil-BWstart-2021}. \hl{Further efficiency is achieved by shifting to the TF$\tau$ space, which localizes wavelet power and reduces the number of subtractions, and by downsampling and heterodyning the initial wavelet transform. Uniquely, MaxWave can render nonwhitened reconstructions, generate \textit{BayesWave}-like error envelopes, and refine its initial reconstruction beyond the fixed TF$\tau$ grid --- all within real-time. Although MaxWave may trade accuracy for speed, refined MaxWave recoveries converge towards those of the robust yet computationally intensive \textit{BayesWave} RJMCMC} \cite{Cornish_2015}, \hl{particularly for higher SNR and mass-ratio transients. Its computational efficiency allows MaxWave to run real-time on multidetector networks, making it a valuable tool for denoising long duration signals and generating large training datasets for glitch classification.}

\section{FUTURE DIRECTIONS AND SCOPE}\label{sec:Future Directions and Scope}

We can apply and extend our algorithm for the rapid, low-latency reconstruction of non-Gaussian features in multiple directions. Our model significantly improves on the existing BayesWave \textit{FastStart} algorithm \cite{Neil-BWstart-2021} that extracts such non-Gaussian features from ground-based detectors. The existing algorithm reduces the convergence time for the \textit{BayesWave} noise model over its large parameter space by staring the sampler's RJMCMC chains near a good initial solution \cite{Neil-BWstart-2021}. Through MaxWave, we can provide a faster and computationally efficient realization of this initial approximate maximum likelihood solution. The central time, central frequency, and time extent of the wavelets we identify can be used as a proposal to the \textit{BayesWave} noise model \cite{Cornish_2015}, encouraging it to put down a wavelet where there is power. To make our proposal legal, it must have a finite proposal density. We can accomplish this by constructing the prior on the amplitude and phase of the proposed wavelet to be a Gaussian blob around the approximate maximum likelihood solution. 

Moreover, we could speed up all likelihood evaluations in the \textit{BayesWave} Glitch model \cite{Cornish_2015} by switching to our modified white wavelets. With analytic wavelet inner products, we could compute an initial wavelet transform for the whitened data and only revisit the data when we occasionally update our noise model. While using the white wavelet basis would lead to a slightly altered signal reconstruction — where the whitened waveform is smooth and the physical waveform has crinkles — we can map the wavelet parameters of the white wavelets and create a reconstruction in the original basis corresponding to the traditional \textit{BayesWave} output.

Our model can facilitate ML algorithms that strive to understand the sources of transient instrument noise. MaxWave can render the time and frequency domain nonwhitened reconstructions of any glitch. This extra information about a glitch’s morphological features can \hl{complement Gravity Spy ML algorithms} \cite{Gravity_Spy_2017}, \hl{which currently only use Qscans to understand instrument noise. The data whitening process and Gaussian background in Qscans can modify the apparent structure of glitches, affecting how ML models interpret them. Although MaxWave’s reconstructions are also altered by the number of wavelets identified and the chosen time-frequency resolution, they can serve as complementary inputs, enhancing ML-based studies for glitch identification and classification.}

We also plan on using our algorithm to denoise slowly varying longer-duration signals, such as the stochastic gravitational wave background and continuous waves from pulsars. We shall extend our analysis to a signal model by approximately maximizing the likelihood across multiple detectors. To do so, we intend to search for a common set of wavelets that can model signals in all detectors while accounting for time and phase shifts and amplitude scaling. We can then deploy the model as a low-latency burst search, which would be an alternative to the existing Coherent WaveBurst algorithm \cite{Klimenko-2005, Klimenko-2016} developed over the last 20 years. 

Once extended to the signal model, our algorithm can offer a real-time, low-latency search for entire LVK observation runs \cite{LVK}, complementing the sophisticated \textit{BayesWave} CBC+Glitch \cite{Sophie-2022} and QuickCBC \cite{Neil_QuickCBC} algorithms. While these methods effectively identify and remove glitches, they are computationally expensive, rely on astrophysical waveform models, and consequently are limited to parameter estimation for specific events \cite{Davis_2022}. In contrast, our approach promises rapid, low-latency, and model-independent glitch subtraction, vastly improving automation in LVK data analysis and equipping us to handle the growing number of events and glitches in future runs.

\section*{ACKNOWLEDGMENTS}
This work was supported by NSF Award No. PHY-2207970. The authors are grateful to Margaret Millhouse and Toral Gupta for help with setting up the \textit{BayesWave} runs.

The authors are grateful for computational resources provided by the LIGO Laboratory and supported by National Science Foundation Grants No. PHY-0757058 and No. PHY-0823459. The data used in our study is based upon work supported by NSF's LIGO Laboratory which is a major facility fully funded by the National Science Foundation.

\bibliography{bibliography_short}

\end{document}